\documentclass{article}

\usepackage{arxiv}

\usepackage[utf8]{inputenc} 
\usepackage[T1]{fontenc}    
\usepackage{hyperref}       
\usepackage{url}            
\usepackage{booktabs}       
\usepackage{amsfonts}       
\usepackage{nicefrac}       
\usepackage{microtype}      
\usepackage{lipsum}		
\usepackage{graphicx}
\usepackage[numbers]{natbib}
\usepackage{doi}

\usepackage{physics}
\usepackage{subfigure}
\usepackage{graphicx}
\usepackage{dcolumn}
\usepackage{bm}
\usepackage{etoolbox} 
\usepackage{ulem}
\usepackage{algorithm2e}
\usepackage{amsfonts,amssymb,amsthm}

\title{Circuit Implementation and Analysis of a Quantum-Walk Based Search Complement Algorithm}

\author{
    \href{https://orcid.org/0000-0003-2344-8505}{\includegraphics[scale=0.06]{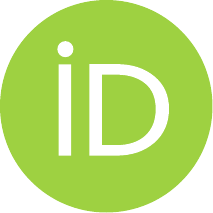}\hspace{1mm}Allan Wing-Bocanegra}\thanks{Email: A00832476@exatec.tec.mx}
    \and
    \href{https://orcid.org/0009-0004-5717-5714}{\includegraphics[scale=0.06]{orcid.pdf}\hspace{1mm}Carlos E. Quintero-Narvaez}\thanks{Email: A00825555@tec.mx}
    \and
    \href{https://orcid.org/0000-0001-7444-4534}{\includegraphics[scale=0.06]{orcid.pdf}\hspace{1mm}Salvador E. Venegas-Andraca}\thanks{Email: venegas@tec.mx}
}

\date{
    Escuela de Ingenieria y Ciencias\\
    Tecnologico de Monterrey\\
    Ave. Eugenio Garza Sada 2501, Monterrey, 64849, N.L., Mexico
}

\renewcommand{\shorttitle}{Circuit Implementation and Analysis of a Quantum-Walk Based Search Complement Algorithm}

\hypersetup{
pdftitle={2405.16322},
pdfsubject={quant-ph, q-bio.QM},
pdfauthor={Allan ~Wing-Bocanegra, Carlos E. ~Quintero-Narvaez, Salvador E. ~Venegas-Andraca},
pdfkeywords={Quantum Walks, Quantum Circuits, Search Algorithm, IBM Quantum, Multigraphs, Qiskit},
}

\begin{document}
\maketitle

\begin{abstract}
We propose a modified version of the quantum walk-based search algorithm created by Shenvi, Kempe and Whaley, also known as the SKW algorithm. In our version of the algorithm, we modified the evolution operator of the system so that it is composed by the product of the shift operator associated to the $2^n$-complete graph with self-loops and a perturbed coin operator based on the Hadamard operator that works as an oracle for the search. The modified evolution operator leads the opposite behavior as in the original algorithm, that is, the probability to measure the target state is reduced. We call this new behavior the \textit{search complement}. Taking a multigraph and matrix approach, we were able to explain that the new algorithm decreases the probability of the target state given that there are less paths that lead towards the node that is associated to the target state in a Unitary Coined Discrete-Time Quantum Walk. The search complement algorithm was executed experimentally on IBM quantum processor \textit{ibmq\_manila} obtaining statistical distances $\ell_1\leq 0.0895$ when decreasing the probability of one state out of four.
\end{abstract}


\section*{Introduction}

A search algorithm is a computational process that is used to locate a specific item or group of items within a larger collection of data. Search algorithms constitute both a solid and widely studied set of computational tools used in scientific and industrial applications everyday (e.g., \cite{knuth97,moore2011,du2016search,bohm_2021,Rahim_2017, zabinsky_2011}), partly due to the emergence of new computer platforms (e.g., mobile and distributed systems), as well as an active research area (e.g., \cite{LIS270,youssef2022}). Concrete examples of fields that make extensive use of search algorithms include Database Management \cite{begg_2019}, Natural Language Processing \cite{wiseman_rush_2016}, Convolutional Neural Networks \cite{poma_melin_gonzález_martínez_2019}, and Computer Networks \cite{saniee_abadeh_habibi_barzegar_sergi_2007}.

Quantum computing has also addressed this problem through different frameworks. The most famous proposal is Grover's algorithm \cite{grover_algorithm_paper}, which is able to solve the search problem with complexity $O(\sqrt{N})$, and was proven to be optimal \cite{zalka_1999}. Another example of a quantum search algorithm is the SKW algorithm \cite{shenvi_kempe_whaley_2003}, proposed by N. Shenvi, J. Kempe and K.B. Whaley, which uses the framework of Unitary Coined Discrete-Time Quantum Walks (UCDTQW).

According to \cite{wbva202301}, a UCDTQW consists of three elements: the state of a quantum walker, $\ket{\psi}$, the evolution operator of the system $U$ and the set of measurement operators of the system $\{M_k\}$. The quantum state of a walker is a bipartite state that is composed of a coin state, $\ket{c_i}$, that is part of an $m$-dimensional Hilbert space $H_C$, and a position state, $\ket{v_j}$, that is part of an $n$-dimensional Hilbert space $H_C$, such that $\ket{\psi}$ is a linear combination of the tensor product between pairs of coin and position states, i.e. $\ket{\psi} = \sum\limits_{i}\sum\limits_{j} a_{ij}\ket{c_i}\otimes\ket{v_j}$. The evolution operator of the system is a bipartite operator that takes the form $U = SC$, where $C$ is the coin operator of the system, which modifies only the coin state in $\ket{\psi}$, i.e. it has the form $C=C'\otimes I_n$, and $S$ is the shift operator of the system, which in principle could by any bipartite operator, and it codifies the information about connections of the graph where the UCDTQW takes place. In general, $S$ is always associated to a directed multigraph. The elements of the set of measurement operators, $\{M_k\}$, have the form $M_k = I_{m}\otimes\ket{v_k}\bra{v_k}$. The SKW algorithms performs a UCDTQW to let a quantum walker move along the vertices of a hypercube graph and search for a marked node.

The SKW algorithm has been studied in detail through theoretical calculations and numerical simulations \cite{potocek_gabris_kiss_jex_2009, zhang_bao_wang_fu_2015, tonchev_2015, douglas_wang_2013, houk_2020}, and, in fact, in \cite{douglas_wang_2013} the complexity of the quantum circuit was reduced by performing the same algorithm, but using the shift operator associated to the $2^n$-dimensional complete graph with self-loops, $\mathcal{K}_{2^n}$, instead of the shift operator associated to a hypercube. However, even after this improvement, the SKW algorithm has never been reported to be efficiently implemented in a general-purpose quantum computer given that the quantum circuit form of the algorithm uses a multi-control Grover operator as part of the coin operator of the UCDTQW, which decomposes into a polynomial number of quantum gates, following the decompositions proposed in \cite{Diao_Zubairy_Chen_2002} and \cite{da_Silva_Park_2022}, making it challenging to run efficiently on NISQ computers.

In view of the former statement, we propose to modify the coin operator of the UCDTQW to make the SKW algorithm more efficient. A natural option arises with the $n$-qubit Hadamard operator, a commonly used operator in the field of UCDTQW for the role of the coin operator, which is indeed less computationally expensive, given that it consists of $n$ sigle-qubit Hadamard gates. Thus one of the purposes of this work is to study the behaviour of the UCDTQW using the Hadamard coin. Moreover, it is also our goal to efficiently implement the search algorithm in IBM's general-purpose quantum computers, thus we decided to take the idea proposed in \cite{douglas_wang_2013} and perform the search algorithm on the graph $\mathcal{K}_{2^n}$. However we use the shift operator proposed in \cite{wbva202302} for the graph $\mathcal{K}_{2^n}$, given that it can be decomposed into a smaller number of \textit{CNOT} gates, which are native gates in IBM's quantum computers.

The combination of the Hadamard operator as a building block for the coin operator and the shift operator associated to the graph $\mathcal{K}_{2^n}$ let to a different behavior than the original SKW algorithm. In this case if we perform a single-step UCDTQW and the measure, the probability of the target state would be reduced with respect to the probabilities of all the other states, which are, in fact, equally distributed. That is to say, the result of this version of the algorithm is the \textit{search complement}.

In \cite{search_complement_grover} a modification to Grover's algorithm was performed to output the search complement too. The search complement algorithm was then used as an initialization subroutine in the QAOA algorithm \cite{qaoa_paper}. However, a UCDTQW version of search complement algorithm had not been exhibited before.

To explain the behavior of the search complement algorithm, we performed a matrix analysis of the evolution operator of the UCDTQW, based on the multigraph theory for UCDTQW developed in \cite{wbva202301, wbva202302}. Furthermore, given that in the UCDTQW model of computation the probability distribution of the quantum walker after $n$ steps is a key component for algorithmic analysis, we introduce a new operator we name as the \textit{probability matrix} of the system. The probability matrix is a matrix that contains the probability distributions associated to all combinations of coin and position states, and is useful to describe our interpretation of the measurement process in the multigraph UCDTQW framework.

Finally, the circuit version of the UCDTQW search complement algorithm was executed through digital simulations using Qiskit and on IBM's quantum computers, providing empirical evidence of their functioning.

\section{\label{DTQW_multigraphs}Introduction to Unitary Coined Discrete-Time Quantum Walks on Multigraphs} 

A Unitary Coined Discrete-Time Quantum Walk (UCDTQW) is composed of three elements: the quantum state of a quantum walker, the evolution operator and the set of measurement operators of the system. The quantum state of a UCDTQW is composed by the coin state and the position state, and in general it can be written as 
\begin{equation}
    |\psi(t)\rangle = \sum\limits_{i,j}a_{ij}|c_i\rangle\otimes|v_j\rangle
    \label{state_of_a_walker}.
\end{equation}
where $|c\rangle$ is called the \textit{coin state} and $|v\rangle$ is the \textit{position state}, each written in terms of the canonical basis of corresponding Hilbert spaces of size $m$ and $n$, respectively. The $n$th step of a UCDTQW is given by

\begin{equation}
    \ket{\psi(t)} = U^n\ket{\psi_0}
\end{equation}
where $\ket{\psi_0}$ is the initial state of the UCDTQW.

If we express the coin states of Eq. \eqref{state_of_a_walker} in explicit matrix notation, we can group all position states with the same coin and rewrite the state of the walker as
\begin{align}
\label{explicit_walker_state}
|\psi(t) \rangle =
\begin{pmatrix}
|V_0(t) \rangle \\
|V_1(t) \rangle \\
\vdots \\
|V_{m-1}(t) \rangle
\end{pmatrix}
\end{align}
That is, $|\psi (t)\rangle$ is an $m$-dimensional block vector, where each entry contains a position subvector $|V_i\rangle$ of size $n$ that is a linear combination of all position states with the same coin state $|c_i\rangle$. As a consequence, whenever an operator is applied on Eq. \eqref{explicit_walker_state}, only $n$ subsequent columns will have an effect on the substate $|V_i(t)\rangle$, which makes it convenient to write operators in block matrix notation, i.e., in general

\begin{equation}
U=
\label{block_unitary}
\begin{pmatrix}
{U}_{00} & {U}_{01} & \dots & U_{0  m-1} \\
{U}_{10} & {U}_{11} & \dots & U_{1  m-1} \\
\vdots & \vdots & \ddots & \vdots \\
{U}_{m-1  0} & {U}_{m-1  1} & \dots & U_{m-1  m-1}
\end{pmatrix}
\end{equation}
where each entry $U_{ij}$ is an $n \times n$ matrix. From Eqs. \eqref{explicit_walker_state} and \eqref{block_unitary}, we can see that when we multiply $U$ and $|\psi (t)\rangle$ only the $i$-th block column of $U$ will affect the $i$-th position subvector of $|\psi (t)\rangle$. 

The fact that an operator acting on a bipartite system can be written in block matrix notation was studied in \cite{petz_1970},  where it was proven that if $U$ is unitary, each column constitutes a set of Kraus operators. That is, the elements of $U$ follow Eq. (\ref{kraus_condition_rows}).


\begin{equation}
    \label{kraus_condition_rows}
    \sum\limits_i U_{ik}^{\dagger}U_{il} = \delta_{kl}I_n
\end{equation}
Given that $U^{\intercal}$ is also unitary it is also true that

\begin{equation}
    \label{kraus_condition_columns}
    \sum\limits_i U_{ki}^{\dagger}U_{li} = \delta_{kl}I_n
\end{equation}

Similar to the transition matrix in a random walk, the evolution operator, $U$, of a UCDTQW is related to a graph $\mathcal{G}$ on which the quantum walker moves from node to node, although in this case we consider $\mathcal{G}$ to be a directed multigraph, as proposed in \cite{wbva202302}. In this context, it is specially convenient that the nodes of $\mathcal{G}$ are labeled with integers so that we can then associate the $i$-th node of $\mathcal{G}$ with the position state $|v_i\rangle$, where $|v_i\rangle$ is the canonical vector with a $1$ in the $i$-th entry and zeros in the rest of the entries. This way, the purpose of each application of $U$ on $|\psi\rangle$ is to make the position state of the walker transition between states, which in turn represents the motion of a walker through the nodes of $\mathcal{G}$.

The evolution operator of a UCDTQW is made by the composition of two operators: the coin operator, $C$, which acts only on the coin states of a quantum walk and puts them into a superposition, and the shift operator, $S$, which is responsible for the transitions in the position states. Thus, $S$ must contain the information of the connections between nodes of $\mathcal{G}$. Given that $S$ is also a block matrix, we can consider each block entry to behave similarly to an independent adjacency matrix that contains only a portion of the connection between vertices in $\mathcal{G}$, in such a way that the addition of all block entries of $S$ yields a matrix with the same information that the adjacency matrix associated to $\mathcal{G}$ \cite{wbva202302}. In other words, to encode the information of the connections between nodes of the graph $\mathcal{G}$ in a shift operator $S$, we must decompose the adjacency matrix $\mathcal{A}$ associated to $\mathcal{G}$ into $m^2$ matrices of size $n \times n$, i.e.,

\begin{equation}
    \label{adjacency_decomposition}
    \mathcal{A} = \sum\limits_{i=0}^{m-1}\sum\limits_{j=0}^{m-1} \mathcal{B}_{ij}
\end{equation}
and then use the matrices $\mathcal{B}_{ij}$ as the blocks of $S$, as follows

\begin{equation}
S=
\label{block_matrix_shift}
\begin{pmatrix}
\mathcal{B}^{\intercal}_{00} & \mathcal{B}^{\intercal}_{01} & \dots & \mathcal{B}^{\intercal}_{0  m-1} \\
\mathcal{B}^{\intercal}_{10} & \mathcal{B}^{\intercal}_{11} & \dots & \mathcal{B}^{\intercal}_{1  m-1} \\
\vdots & \vdots & \ddots & \vdots \\
\mathcal{B}^{\intercal}_{m-1  0} & \mathcal{B}^{\intercal}_{m-1  1} & \dots & \mathcal{B}^{\intercal}_{m-1  m-1}
\end{pmatrix}
\end{equation}
With the condition that Eqs. \eqref{kraus_condition_rows} and \eqref{kraus_condition_columns} must be satisfied simultaneously, i.e., each block column and block row of $S$ forms a set of Kraus operators. Notice that the block elements of $S$ are transposed, this is due to the fact that if and we apply directly the adjacency matrix of some directed graph to some vector of the canonical basis that represents the position of a walker, then the result of that operation will output a new vector corresponding to the walker moving in the opposite direction to the arcs of $\mathcal{G}$. If we apply the transpose matrix, the output vector will correspond to the walker moving in the same direction of the arcs of $\mathcal{G}$. Given that the coin and position states associated to a quantum walker are associated to canonical vectors, the same logic applies to $S$.

A special case of Eqs.~\eqref{kraus_condition_rows} - \eqref{block_matrix_shift} happens when $S$ is composed of unitary matrices placed in the main diagonal \cite{montanaro_2007, godsil_zhan_2019, wbva202301}, and the rest of the entries are the zero matrix, as shown in Eq. \eqref{block_diag_shift}.

\begin{equation}
S =
\label{block_diag_shift}
\begin{pmatrix}
\mathcal{B}^{\intercal}_0 & 0 & \dots & 0 \\
0 & \mathcal{B}^{\intercal}_1 & \dots & 0 \\
\vdots & \vdots & \ddots & \vdots \\
0 & 0 & \dots & \mathcal{B}^{\intercal}_{m-1} 
\end{pmatrix}
\end{equation}


For example, consider a UCDTQW on a complete graph with self-loops, that is, a graph in which all nodes are directly connected with one another and with themselves, thus the adjacency matrix of a complete graph is given by an $n\times n$ matrix full of ones. This quantum walk is represented by the symbol $\mathcal{K}_{2^n}$, where $2^n$ indicates the number of nodes. In this case let $n=2$. According to Eq. \eqref{adjacency_decomposition}, to map the adjacency matrix $\mathcal{A}$ of a graph $\mathcal{G}$ into the shift operator of a DTQW, we must decompose $\mathcal{A}$ into a sum of matrices each associated to a subgraph of $\mathcal{G}$. Then, the following decomposition is possible

\begin{multline}
\mathcal{A} =
\begin{pmatrix}
    1 & 1 & 1 & 1 \\
    1 & 1 & 1 & 1 \\
    1 & 1 & 1 & 1 \\
    1 & 1 & 1 & 1
    \end{pmatrix}=
\begin{pmatrix}
    1 & 0 & 0 & 0 \\
    0 & 0 & 0 & 0 \\
    0 & 0 & 0 & 0 \\
    0 & 0 & 0 & 0
    \end{pmatrix}+
\begin{pmatrix}
    0 & 1 & 0 & 0 \\
    0 & 0 & 0 & 0 \\
    0 & 0 & 0 & 0 \\
    0 & 0 & 0 & 0
    \end{pmatrix}+
\begin{pmatrix}
    0 & 0 & 1 & 0 \\
    0 & 0 & 0 & 0 \\
    0 & 0 & 0 & 0 \\
    0 & 0 & 0 & 0
    \end{pmatrix}+
\begin{pmatrix}
    0 & 0 & 0 & 1 \\
    0 & 0 & 0 & 0 \\
    0 & 0 & 0 & 0 \\
    0 & 0 & 0 & 0
    \end{pmatrix}\\ \hspace{4.9em} +
\begin{pmatrix} 
    0 & 0 & 0 & 0 \\
    1 & 0 & 0 & 0 \\
    0 & 0 & 0 & 0 \\
    0 & 0 & 0 & 0
    \end{pmatrix}+
\begin{pmatrix}
    0 & 0 & 0 & 0 \\
    0 & 1 & 0 & 0 \\
    0 & 0 & 0 & 0 \\
    0 & 0 & 0 & 0
    \end{pmatrix}+
\begin{pmatrix}
    0 & 0 & 0 & 0 \\
    0 & 0 & 1 & 0 \\
    0 & 0 & 0 & 0 \\
    0 & 0 & 0 & 0
    \end{pmatrix}+
\begin{pmatrix}
    0 & 0 & 0 & 0 \\
    0 & 0 & 0 & 1 \\
    0 & 0 & 0 & 0 \\
    0 & 0 & 0 & 0
    \end{pmatrix}\\ \hspace{4.9em} +
\begin{pmatrix}
    0 & 0 & 0 & 0 \\
    0 & 0 & 0 & 0 \\
    1 & 0 & 0 & 0 \\
    0 & 0 & 0 & 0
    \end{pmatrix}+
\begin{pmatrix}
    0 & 0 & 0 & 0 \\
    0 & 0 & 0 & 0 \\
    0 & 1 & 0 & 0 \\
    0 & 0 & 0 & 0
    \end{pmatrix}+
\begin{pmatrix}
    0 & 0 & 0 & 0 \\
    0 & 0 & 0 & 0 \\
    0 & 0 & 1 & 0 \\
    0 & 0 & 0 & 0
    \end{pmatrix}+
\begin{pmatrix}
    0 & 0 & 0 & 0 \\
    0 & 0 & 0 & 0 \\
    0 & 0 & 0 & 1 \\
    0 & 0 & 0 & 0
    \end{pmatrix}\\ \hspace{4.9em} +
\begin{pmatrix}
    0 & 0 & 0 & 0 \\
    0 & 0 & 0 & 0 \\
    0 & 0 & 0 & 0 \\
    1 & 0 & 0 & 0
    \end{pmatrix}+
\begin{pmatrix}
    0 & 0 & 0 & 0 \\
    0 & 0 & 0 & 0 \\
    0 & 0 & 0 & 0 \\
    0 & 1 & 0 & 0
    \end{pmatrix}+
\begin{pmatrix}
    0 & 0 & 0 & 0 \\
    0 & 0 & 0 & 0 \\
    0 & 0 & 0 & 0 \\
    0 & 0 & 1 & 0
    \end{pmatrix}+
\begin{pmatrix}
    0 & 0 & 0 & 0 \\
    0 & 0 & 0 & 0 \\
    0 & 0 & 0 & 0 \\
    0 & 0 & 0 & 1
    \end{pmatrix}\\   \hspace{9.15em}
\label{k4_adj_whole_decomposition}
\end{multline}

\begin{figure}[h!]
\subfigure[]{
\centering
\resizebox{0.48\textwidth}{!}{ 
\includegraphics[scale=0.45]{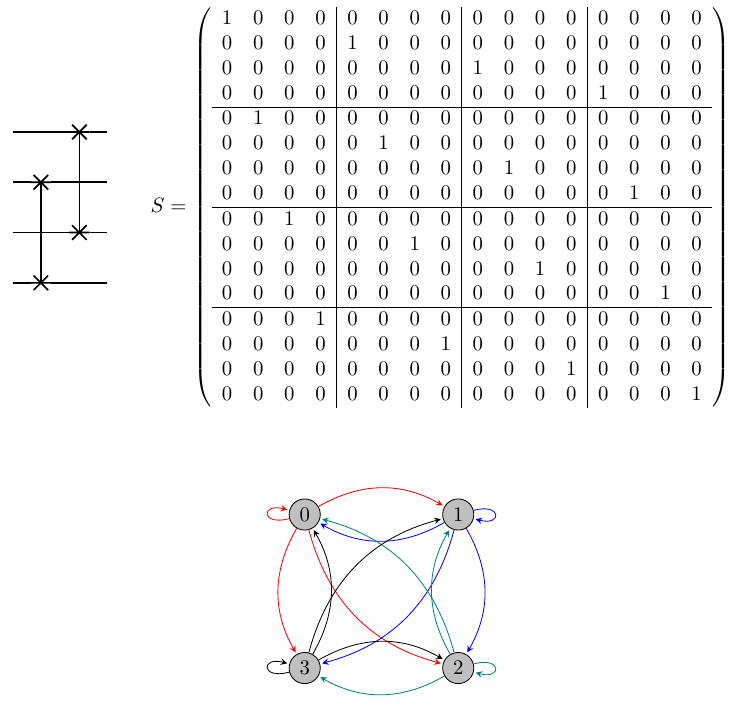}
}
\label{model_circuit_example}
}
\subfigure[]{
\centering   
\resizebox{0.48\textwidth}{!}{ 
\includegraphics[scale=0.45]{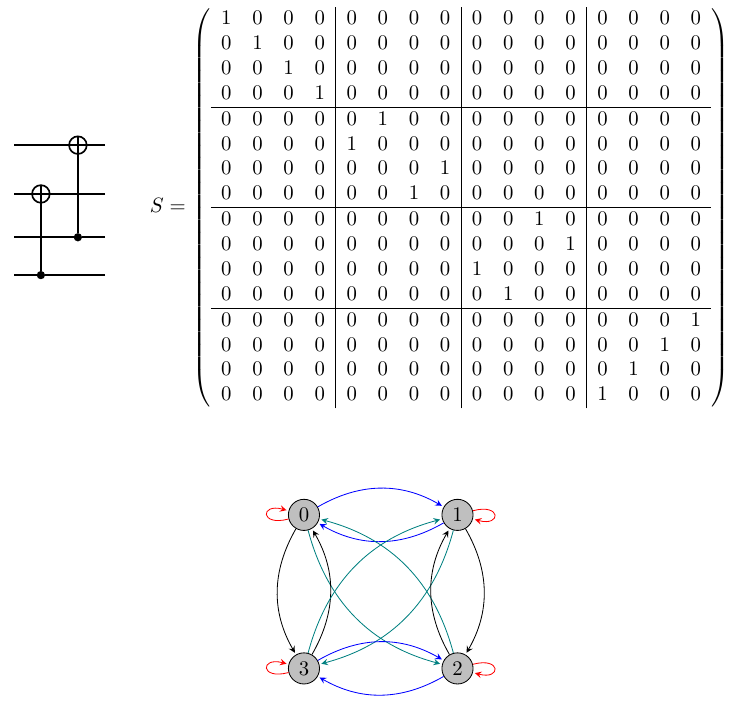}
}
\label{model_circuit_corollary_example}
}
\caption{Figure extracted from \cite{wbva202302}. Both figures show the quantum circuit, unitary matrix and multigraph representation associated to the graph $\mathcal{K}_4$. In both circuits we consider the top two qubits correspond to the position register and the bottom two qubits correspond to the coin state. As a consequence of having a bipartite system, the unitary matrix representation of the circuit is split into $4 \times 4$ blocks. Each column of the matrices is associated to a color in the multigraph representation of $\mathcal{K}_4$: the first, second, third and fourth columns are associated to red, blue, green and black, respectively. (a) and (b) present the quantum circuits resulting from mapping eq. \eqref{k4_adj_whole_decomposition} and eq. \eqref{k4_adj_shunt_decomposition} into a unitary matrix, respectively.}
\label{model_circuit_example_both_cases}
\end{figure}

Matrices in Eq. \eqref{k4_adj_whole_decomposition} were used as blocks to compose the shift operator associated to a UCDTQW on $\mathcal{K}_4$ in \cite{douglas_wang_2009}. Fig. \ref{model_circuit_example} displays the shift operator, the quantum circuit and the graph representation of $\mathcal{K}_4$ according to \cite{douglas_wang_2009}.

In \cite{wbva202301} an alternative decomposition for the adjacency matrix of $\mathcal{K}_{2^n}$ based on eq. \eqref{block_diag_shift} was proposed. Using the decomposition proposed in \cite{wbva202301}, we can rewrite $\mathcal{A}$ as

\begin{multline}
\mathcal{A} =
\begin{pmatrix}
    1 & 1 & 1 & 1 \\
    1 & 1 & 1 & 1 \\
    1 & 1 & 1 & 1 \\
    1 & 1 & 1 & 1
    \end{pmatrix}=
\begin{pmatrix}
    1 & 0 & 0 & 0 \\
    0 & 1 & 0 & 0 \\
    0 & 0 & 1 & 0 \\
    0 & 0 & 0 & 1
    \end{pmatrix}+
\begin{pmatrix}
    0 & 1 & 0 & 0 \\
    1 & 0 & 0 & 0 \\
    0 & 0 & 0 & 1 \\
    0 & 0 & 1 & 0
    \end{pmatrix}+
\begin{pmatrix}
    0 & 0 & 1 & 0 \\
    0 & 0 & 0 & 1 \\
    1 & 0 & 0 & 0 \\
    0 & 1 & 0 & 0
    \end{pmatrix}+
\begin{pmatrix}
    0 & 0 & 0 & 1 \\
    0 & 0 & 1 & 0 \\
    0 & 1 & 0 & 0 \\
    1 & 0 & 0 & 0
    \end{pmatrix}   
\label{k4_adj_shunt_decomposition}    
\end{multline}
Fig. \ref{model_circuit_corollary_example}, displays the shift operator, quantum circuit and graph representation associated to $\mathcal{K}_4$ using the decomposition in Eq. \eqref{k4_adj_shunt_decomposition}. 

In both Figs. \ref{model_circuit_example}  and  \ref{model_circuit_corollary_example}, the operators are partitioned into $4 \times 4$ block matrices, and each color of arcs of the graphs below them corresponds to a different column, i.e., red, blue, green and back arcs are associated to the first, second, third and fourth columns, respectively. Notice that although the structure is the same, the colors of the arcs vary. This indicates that when each representation of $S$ is applied to the same quantum state of a walker $|\psi\rangle$, the walker will move differently from one graph to another. This in turn might lead to a different behaviour when each representation of $S$ is used to run quantum walk-based algorithms, as we will present in the next sections.

To complete the evolution operator, we must now describe the coin operator of the system. The coin operator of a UCDTQW is an operator that acts only on the coin register and has the action to put all coin states $|c_i\rangle$ of the system in a superposition of states, leaving the position register intact. It is usually defined as $C \otimes I_n$.

\begin{figure}
    \centering
\resizebox{0.65\textwidth}{!}{    
\includegraphics{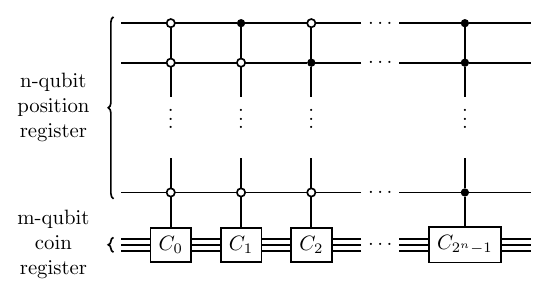}
}
    \caption{Figure extracted from \cite{wbva202302}. Consider black controls to be 1's and white controls to be 0's. Thus, every coin operator $C_i$ is activated when the position register has a specific state. This property will be useful to perform quantum walk-based algorithms in which a specific condition is met when a walker steps onto a specific node.}
    \label{general_coin_ciq}
\end{figure}

The coin operator must not modify the states of the position register, although we can still make it dependent on the position of the walker. We consider the generalization of the coin operator proposed in \cite{wbva202302}

\begin{equation}
    \label{general_coin}
    \mathcal{C}=\sum\limits_{k=0}^{n-1} C_{k} \otimes |v_k\rangle \langle v_k|
\end{equation}
This generalization allows $\mathcal{C}$ to act differently depending on the position state $|v_k\rangle$, and that property is key for running the SKW algorithm. In circuit form, Eq. \eqref{general_coin} can be defined as presented in Fig. \ref{general_coin_ciq}, where we show a sequence of fully controlled gates, with a different pattern of white and black controls. We must interpret this pattern of controls as binary code, in such a way that black controls represent a 1, white controls represent a 0, and the uppermost control corresponds to the least significant bit. This way, each coin operator $C_i$ will act only on the position state whose binary representation coincides with the sequence of black and white dots that controls it. The mapping to binary of a state represented by an integer can be done as explained in \cite{wbva202301}.

Next, we define the measurement operator of position state $|v_j\rangle$ as

\begin{equation}
     M_j = I_m\otimes|v_j\rangle\langle v_j|
\label{measurement_operator}
\end{equation}
Thus, the probability of finding a walker on node $v_j$ is given by

\begin{equation}
    p(v_j) = \langle \psi(t)|M^{\dagger}_j M_j| \psi(t) \rangle
    \label{node_probability_eq}
\end{equation}
If we let $v_{ij}$ be the entry of $|\psi(t)\rangle$ associated to subvector $i$ and node $j$, then $p(v_j)$ can be written as

\begin{equation}
    p(v_j)=\sum\limits_i v^{*}_{ij}v_{ij} = |v_{ij}|^2
    \label{single_node_probability}
\end{equation}
This allows us to create a probability distribution vector as shown in Eq. \eqref{probability_vector_measurement}.

\begin{equation}
    P= 
\begin{pmatrix}
p(0)\\
p(1)\\
\vdots\\
p(n-1)
\end{pmatrix}=\begin{pmatrix}
\sum\limits_i |v_{i0}|^2\\
\sum\limits_i |v_{i1}|^2\\
\vdots\\
\sum\limits_i |v_{in-1}|^2
\end{pmatrix}
\label{probability_vector_measurement}
\end{equation}
Notice that Eq. \eqref{probability_vector_measurement} is equivalent to Eq. \eqref{probability_vector_hadamard}

\begin{equation}
P=\sum\limits_i(|V_i\rangle)^{*}\odot|V_i\rangle    
\label{probability_vector_hadamard}
\end{equation}
where $|V_i\rangle$ is the $i$th position subvector in $|\psi_0\rangle$, as shown in eq. \eqref{explicit_walker_state}, and $\odot$ represents the element-wise product, also known as Hadamard product.

Eqs. \eqref{probability_vector_measurement} and \eqref{probability_vector_hadamard} present a compact and easy way to calculate the probability distribution for all nodes in a graph where the UCDTQW takes place. Furthermore, based on Eq. \eqref{probability_vector_hadamard} we can obtain a matrix with the probability distributions associated to all initial states of the form $\ket{\psi_0} = \ket{c_i}\otimes\ket{v_j}$ where $\ket{c_i}$ and $\ket{v_j}$ are associated to canonical vectors in the Hilbert space where they belong. If the coin and position states of a UCDTQW are standard basis vectors, then the initial quantum state of a walker, $|\psi_0\rangle=|c_i\rangle\otimes|v_j\rangle$, is always a standard basis vector too, that is, $|\psi_0\rangle$ is a vector with a 1 in the $k$th entry, and zeroes in the rest of the entries. Then, if we apply an evolution operator $U^n$ to $|\psi_0\rangle$, the result of this operation will be a vector that contains the entries of the $k$th column of $U^n$. This resulting vector will be the quantum state of a walker after $n$ steps, and the probability distribution of the walker after $n$ steps can be calculated with Eq. \eqref{probability_vector_hadamard}. If we vary the states $\ket{c_i}$ and $\ket{v_j}$ that compose $\ket{\psi_0}$, then we are able to obtain the probability distributions of all combinations of coin and position states applying Eq.  \eqref{probability_vector_hadamard}.

We can use the same rationale for all the columns of $U^n$ to calculate at once the probability distributions corresponding to all the possible initial states of a quantum walk that is evolved using $U$. To do so, we perform the entry-wise product of the whole evolution operator $U^n$ with its complex conjugate, that is, we define $V = (U^n)^{*}\odot(U^n)$. Then, the matrix containing the probability distributions of all the possible initial states of a quantum walk is obtained by adding up all the block rows of $V$. We call the result, the \textit{probability matrix} of the system, which we denote by $M_P$.

To provide an explicit explanation of the former statement recall that $U$ is a block matrix composed of submatrices $U_{ij}$ (see Eq. \eqref{block_matrix_shift}). Then, if we consider the entries of $U$ to be $v_{ij}^{kl}$, the entry-wise product between $(U^n)^{*}$ and $U^n$ gives out Eq. \eqref{hadamard_product_U}. Notice that the subindices of $v_{ij}^{kl}$ indicate to which submatrix, $U_{ij}$, $v_{ij}^{kl}$ belongs too, while the superindices indicate the position of $v_{ij}^{kl}$ in $U_{ij}$.

\begin{figure}[h!]
\centering
\resizebox{0.85\textwidth}{!}{  
\includegraphics{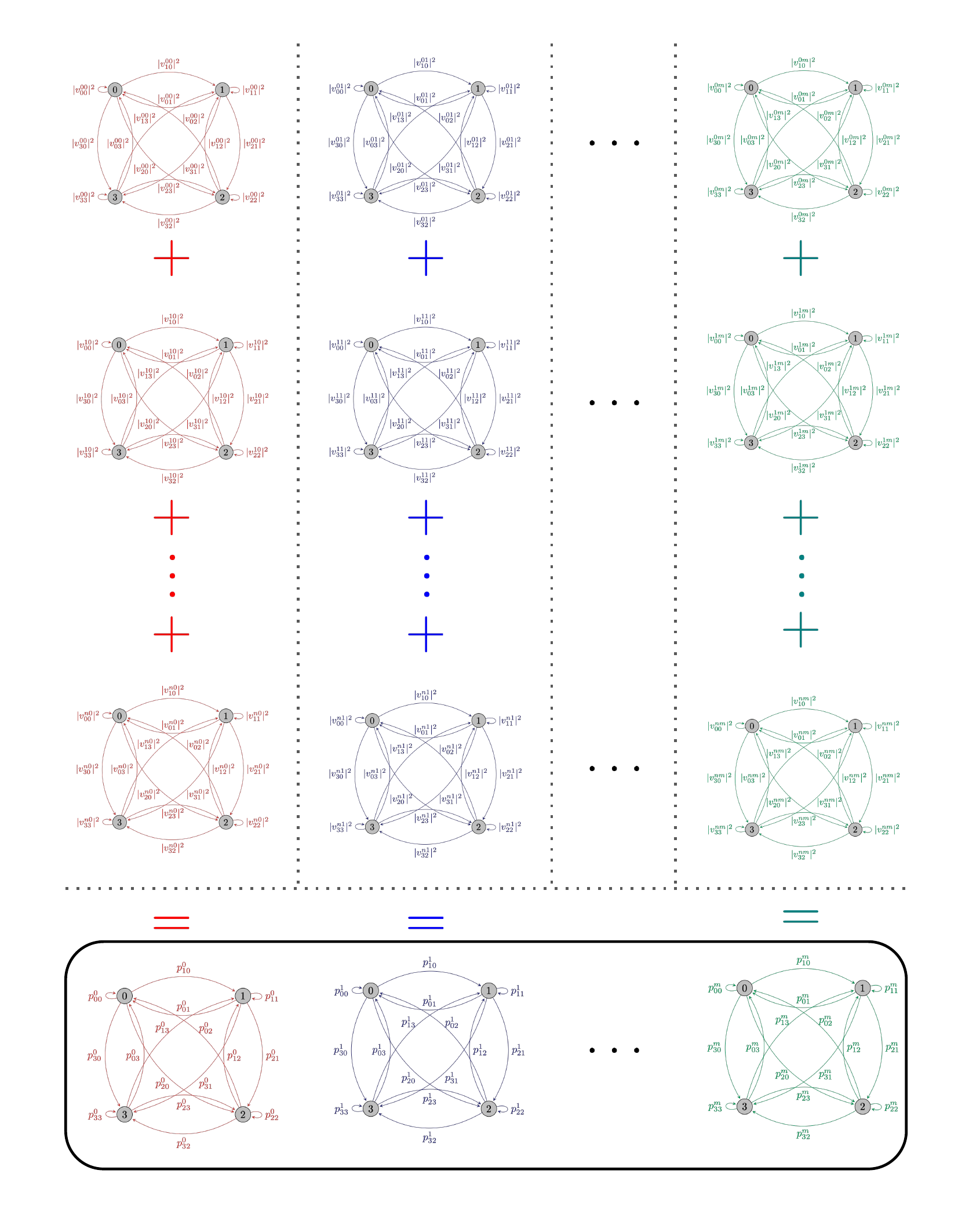}
}
\caption{Visual representation of the transformation process of the directed multigraph associated to a bipartite unitary operator, $U^n$, in a UCDTQW when the position register is measured. The upper part of the figure corresponds to the multigraph representation of a special case of the operator $V = (U^n)^{*}\odot (U^n)$ (see \eqref{hadamard_product_U}) where each matrix has $4$ entries. In this part of the figure, each subset of graphs separated by a dotted line is associated to a different coin state. The lower part of the figure represents the probability matrix associated to $V$, where $p_{qr}^{l} = \sum\limits_{i}|v_{qr}^{il}|^2$. The complete figure represents how the arcs that connect vertex $i$ with vertex $j$ in the upper part of the figure are transformed into one single arc whose weight is the sum of the weights all arcs in the upper part of the figure. Furthermore, weight of the resulting arc is a probability.}
\label{multigraph_arc_collapse}
\end{figure}

\begin{equation}
\resizebox{0.9\textwidth}{!}{  
$  V = (U^n)^{*}\odot(U^n) =
    \left(
    \arraycolsep=2pt\def\arraystretch{1}
    \begin{array}{ *{5}{c} | *{3}{c} | *{5}{c}}
    |v_{00}^{00}|^2 & |v_{01}^{00}|^2 & \dots & |v_{0(n-1)}^{00}|^2 & |v_{0n}^{00}|^2 &  & \dots &  & |v_{00}^{0m}|^2 & |v_{01}^{0m}|^2 & \dots & |v_{0(n-1)}^{0m}|^2 & |v_{0n}^{0m}|^2 \\
    |v_{10}^{00}|^2 & \ddots & \ddots & \vdots & \vdots &  & \dots &  & |v_{10}^{0m}|^2 & \ddots & \ddots & \vdots & \vdots \\ 
    \vdots & \ddots & |v_{ij}^{00}|^2 & \ddots & \vdots &  & \dots &  & \vdots & \ddots & |v_{ij}^{0m}|^2 & \ddots & \vdots \\
    |v_{(n-1)0}^{00}|^2 & \dots & \ddots & \ddots & |v_{(n-1)n}^{00}|^2 &  & \dots &  & |v_{(n-1)0}^{0m}|^2 & \dots & \ddots & \ddots & |v_{(n-1)n}^{0m}|^2 \\
    |v_{n0}^{00}|^2 & \dots & \dots & |v_{n(n-1)}^{00}|^2 & |v_{nn}^{00}|^2 &  & \dots &  & |v_{n0}^{0m}|^2 & \dots & \dots & |v_{n(n-1)}^{0m}|^2 & |v_{nn}^{0m}|^2 \\ \hline 
     &  &  &  &  &  &  &  &  &  &  &  & \\
    \vdots & \vdots & \vdots & \vdots & \vdots &  & \ddots &  & \vdots & \vdots & \vdots & \vdots & \vdots \\
    &  &  &  &  &  &  &  &  &  &  &  & \\ \hline
    |v_{00}^{m0}|^2 & |v_{01}^{m0}|^2 & \dots & |v_{0(n-1)}^{m0}|^2 & |v_{0n}^{m0}|^2 &  & \dots &  & |v_{00}^{mm}|^2 & |v_{01}^{mm}|^2 & \dots & |v_{0(n-1)}^{mm}|^2 &  |v_{0n}^{mm}|^2 \\
    |v_{10}^{m0}|^2 & \ddots & \ddots & \vdots & \vdots &  & \dots &  & |v_{10}^{mm}|^2 & \ddots & \ddots & \vdots & \vdots \\ 
    \vdots & \ddots & |v_{ij}^{m0}|^2 & \ddots & \vdots &  & \dots &  & \vdots & \ddots & |v_{ij}^{mm}|^2 & \ddots & \vdots \\
    |v_{(n-1)0}^{m0}|^2 & \dots & \ddots & \ddots & |v_{(n-1)n}^{m0}|^2 &  & \dots &  & |v_{(n-1)0}^{mm}|^2 & \dots & \ddots & \ddots & |v_{(n-1)n}^{mm}|^2 \\
    |v_{n0}^{m0}|^2 & \dots & \dots & |v_{n(n-1)}^{m0}|^2 & |v_{nn}^{m0}|^2 &  & \dots &  & |v_{n0}^{mm}|^2 & \dots & \dots & |v_{n(n-1)}^{mm}|^2 & |v_{nn}^{mm}|^2 \\ 
   \end{array}
   \right)$
}   
\label{hadamard_product_U}
\end{equation}

Next, if we add up all block rows of Eq. \eqref{hadamard_product_U}, we obtain the explicit form of $M_P$, as shown in Eq. \eqref{general_probability_matrix}.

\begin{equation}
\resizebox{0.8\textwidth}{!}{  
$   M_P=
    \left(
    \arraycolsep=2pt\def\arraystretch{1}
    \begin{array}{ *{4}{c} | *{3}{c} | *{4}{c} }
    \sum\limits_{i}|v_{00}^{i0}|^2 & \sum\limits_{i}|v_{01}^{i0}|^2 & \hdots & \sum\limits_{i}|v_{0n}^{i0}|^2 &
    &\hdots& & \sum\limits_{i}|v_{00}^{im}|^2 & \sum\limits_{i}|v_{01}^{im}|^2 &\hdots & \sum\limits_{i}|v_{0n}^{im}|^2 \\
    \sum\limits_{i}|v_{10}^{i0}|^2 & \ddots & \ddots & \vdots &
    &\hdots& & \sum\limits_{i}|v_{10}^{im}|^2 & \ddots & \ddots & \vdots\\
    \vdots & \ddots &  \sum\limits_{i}|v_{qr}^{i0}|^2 & \vdots & &\hdots& & \vdots & \ddots &  \sum\limits_{i}|v_{qr}^{i0}|^2 & \vdots \\
    \sum\limits_{i}|v_{(n-1)0}^{i0}|^2 & \hdots & \ddots & \vdots & &\hdots& & \sum\limits_{i}|v_{(n-1)0}^{im}|^2 & \hdots & \ddots & \vdots\\
    \sum\limits_{i}|v_{n0}^{i0}|^2 & \hdots & \hdots & \sum\limits_{i}|v_{nn}^{i0}|^2 & &\hdots& & \sum\limits_{i}|v_{n0}^{im}|^2 & \hdots & \hdots & \sum\limits_{i}|v_{nn}^{im}|^2 \\ 
   \end{array}
   \right)$
}   
   \label{general_probability_matrix}
\end{equation}

From the multigraph framework the submatrices in $V$ and $M_P$ also are associated to directed graphs which conform a directed multigraph. Fig. \ref{multigraph_arc_collapse} shows a visual representation of the transformation suffered by a 4-vertex directed multigraph when $V$ is transformed into $M_P$. The complete figure represents how the arcs that connect vertex $i$ with vertex $j$ in the upper part of the figure merge in a single arc whose weight is the transition probability from vertex $i$ with vertex $j$ after $n$ steps. Thus, we call this transformation the collapse of the multigraph associated to $U^n$.

\section{\label{SKW_alg_section}Quantum Walk Version of a Search Algorithm}

In \cite{shenvi_kempe_whaley_2003}, Shenvi, Kempe and Whaley presented the first UCDTQW-based search algorithm. The proposed algorithm works by performing a UCDTQW on a hypercube, in such a way that, after $\mathcal{O}(2^{n/2})$ steps, the probability of measuring the quantum state associated to the target node will be increased. The evolution operator proposed in \cite{Moore_Russell_2002, Kempe_2003} for the UCDTQW on the hypercube was used in \cite{shenvi_kempe_whaley_2003} to create the search algorithm, and the key element of the algorithm was to induce a perturbation to the coin operator such that it acts differently only on the quantum state associated to the target node of the search. The mathematical expression proposed in \cite{shenvi_kempe_whaley_2003} for the coin operator is 

\begin{equation}
    \mathcal{C}'=C_0\otimes I+(C_1-C_0)\otimes|v_t\rangle\langle v_t|
\label{perturbed_coin_eq}
\end{equation}
where $C_0$ is the original coin, $C_1$ is the perturbation, and $v_t$ is the target node. 

The authors of \cite{shenvi_kempe_whaley_2003} analyzed the perturbed evolution operator of the system, $U'=S\mathcal{C}'$, for the case in which $C_0$ is the Grover operator, $G$, and $C_1=-I_m$, and proved that the evolution of the initial state of the walker, $|\psi_0\rangle$, approximates the target state $|v_t\rangle$ with probability $P(|v_t\rangle)=1/2-O(1/n)$ after $U'$ is applied $\frac{\pi}{2}\sqrt{2^{n}}$ times, with $n$ being the dimension of the hypercube on which the UCDTQW takes place. The search algorithm presented in \cite{shenvi_kempe_whaley_2003} is summarized in Algorithm \ref{quantum_walk_search_algorithm}.

\begin{algorithm}
\caption{Search algorithm proposed in \cite{shenvi_kempe_whaley_2003}}\label{quantum_walk_search_algorithm}\begin{enumerate}
    \item Initialize all qubits of both coin and position registers with state $\ket{0}$ and then apply a Hadamard gate to all qubits of both registers to have superposition of all possible states.
    \item Apply the perturbed evolution operator, $U'=S\mathcal{C}'$, $\frac{\pi}{2} \sqrt{2^{n}}$ times.
    \item Measure the position register in the $|c,v\rangle$ basis.
\end{enumerate}
\end{algorithm}

Given that one of our goals in this work is to efficiently implement the SKW search algorithm on IBM's quantum computers, and in view that the main element of the algorithm is the perturbed coin operator, in the next  we will devote next subsection to the study of the quantum circuit implementation of $\mathcal{C}'$.

\subsection{\label{section_coin_op}Circuit Implementation of the Perturbed Coin Operator}

Let us start by studying the action of the perturbed coin operator on both target and non-target states. Applying $\mathcal{C}'$ to the target state, $\ket{\psi} = \ket{c}\otimes\ket{v_t}$ we get

\begin{align*}
\begin{split}
\mathcal{C}'|c\rangle \otimes |v_t\rangle & =(C_0\otimes I_n + (C_1-C_0)\otimes|v_t\rangle\langle v_t|)|c\rangle\otimes|v_t\rangle \\
 & =C_0\otimes I_n|c\rangle\otimes|v_t\rangle + \left[(C_1-C_0)\otimes|v_t\rangle\langle v_t|\right]|c\rangle\otimes|v_t\rangle\\
 & =C_0 |c\rangle \otimes I_n|v_t\rangle + C_1 |c\rangle \otimes |v_t\rangle - C_0|c\rangle\ \otimes |v_t\rangle\\
 & =C_1|c\rangle\otimes|v_t\rangle \\
\end{split}
\end{align*}
Now, if $C_1 =-I_m$, then
\begin{align*}
 \mathcal{C}'|c\rangle\otimes|v_t\rangle & = -|c\rangle\otimes|v_t\rangle \hspace{13.25em}
\end{align*}
For the case of a non-target state $|v_n\rangle$ we get

\begin{align*}
\begin{split}
\mathcal{C}'|c\rangle\otimes|v_n\rangle & =(C_0\otimes I_n + (C_1-C_0)\otimes|v_t\rangle\langle v_t|)|c\rangle\otimes|v_n\rangle \\
 & = (C_0\otimes I_n)|c\rangle\otimes|v_n\rangle\\
& = C_0|c\rangle\otimes|v_n\rangle
\end{split}
\end{align*}
That is, the action of $\mathcal{C}'$ on the quantum state of a walker, $|\psi\rangle$, is to shift its phase if the position state of $|\psi\rangle$ is the target state, and to put the coin state into a superposition of states, if the position state of $|\psi\rangle$ is not the target state. From Eq. \eqref{perturbed_coin_eq}, notice that the perturbed coin operator can be written as

\begin{equation}
    \mathcal{C}' = \sum\limits_{j\neq t} C_0\otimes|v_j\rangle\langle v_j| + C_1\otimes|v_t\rangle\langle v_t|
\end{equation}
which is just a special case of the general coin operator presented in Eq.~\eqref{general_coin}, which means we can use fully controlled gates to build the perturbed coin operator, as shown in Fig. \ref{general_coin_ciq}.

\begin{figure}[h!]
\centering
\includegraphics[scale=1.1]{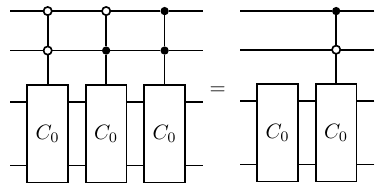}
\caption{Circuit implementation of a perturbed coin operator that enables the search of the state $|1\rangle$ ($|01\rangle$ in two-bit binary representation) in Algorithm \ref{quantum_walk_search_algorithm}.
}
\label{perturbed_coin_circuit}
\end{figure}

To exemplify how to construct the circuit of the perturbed coin operator, suppose that the target state is $|1\rangle$ and that we work with a system where both position and coin registers have two qubits. As a first step, we must rewrite the target state in binary notation using as many bits as the number of qubits of the position register, in this case $|1\rangle \rightarrow |01\rangle$. The same must be done for the rest of position states in this system, i.e. $|0\rangle \rightarrow |00\rangle$, $|2\rangle \rightarrow |10\rangle$ and $|3\rangle \rightarrow |11\rangle$. According to the description of the perturbed coin operator, we want an operator such that when the position state of the walker is $|00\rangle$, $|10\rangle$ or $|11\rangle$, $\mathcal{C}'$ acts as $C_0$ on the coin register, thus $\mathcal{C}'$ will be conformed of three fully controlled $C_0$ gates that are individually activated whenever the position states are $|00\rangle$, $|10\rangle$ or $|11\rangle$. The circuit that represents the perturbed coin operator is shown in the left-hand side of Fig. \ref{perturbed_coin_circuit}. On the right hand side of Fig. \ref{perturbed_coin_circuit}, we find an equivalent circuit which is less expensive to implement and has a controlled gate that acts directly on the target state, $|01\rangle$. This pattern can be generalized for higher-dimensional systems.

Notice that in this circuit we do not take into account the phase shift presented in \cite{shenvi_kempe_whaley_2003} when $\mathcal{C}'$ acts on the state associated to the target node, given that it just represents a global phase which is irrelevant when measurement is performed, and experimentally, it is expensive to implement in quantum circuit form. That is to say, our version of the perturbed coin operator uses $C_1 = I_m$, instead of $C_1 = -I_m$, as proposed in \cite{shenvi_kempe_whaley_2003}.

\subsection{\label{section_whole_search_circuit}Circuit Implementation of the Whole Evolution Operator}

Douglas and Wang \cite{douglas_wang_2013} studied the quantum circuit implementation of the search algorithm on a hypercube, and arrived to the same circuit form of the perturbed coin operator presented in the right-hand side of Fig. \ref{perturbed_coin_circuit}. They used the perturbed Grover coin, as stated in the original algorithm \cite{shenvi_kempe_whaley_2003}, and combining it with the shift operator of the hypercube \cite{Moore_Russell_2002, embedded_hypercubes_2014}, they completed the perturbed evolution operator of the search algorithm.

However, as one of our goals is to experimentally implement the search algorithm on IBM's NISQ computers, we cannot make use of the hypercube, in view that, recently, Wing and Andraca \cite{wbva202301} studied the experimental implementation of the quantum circuit form of the evolution operators associated to the UCDTQW on the line, cycle, hypercube, and complete graphs, and they found out that the quantum circuit associated to the hypercube is computationally expensive, and does not yield efficient experimental results on IBM's NISQ computers, given that it need multi-control \textit{CNOT} gates for its implementation and multi-control \textit{CNOT} decompose into an exponential number of elementary gates. 

\begin{figure}[b!]
\centering
    \centering
\resizebox{0.5\textwidth}{!}{ 
\includegraphics[scale=1.1]{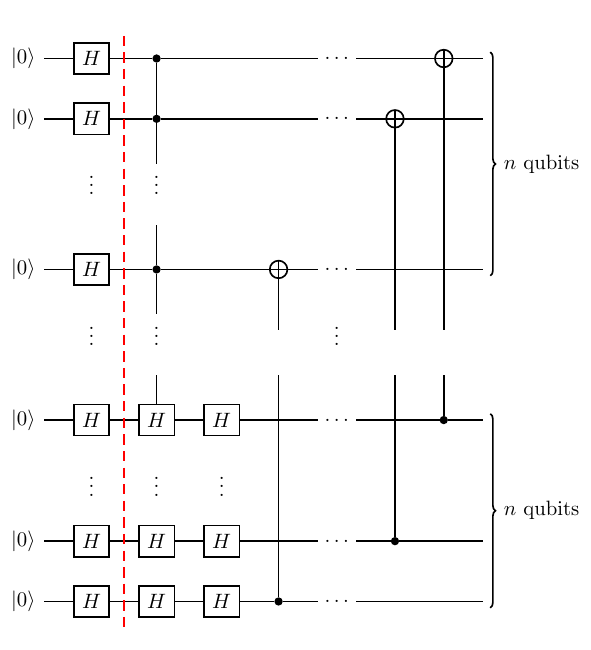}
}
\caption{
General circuit of the search complement of node $2^n-1$ on $\mathcal{K}_{2^n}$. To the left of the dotted line we present the qubits' initialization routine. To the right of the dotted line we present the perturbed evolution operator, $U'$, to perform the search complement of node $2^n-1$ on the graph $\mathcal{K}_{2^n}$. The top $n$ qubits correspond to the position register and the bottom $n$ qubits correspond to the coin register
}
\label{general_circuit_search_complement}
\end{figure}

Douglas and Wang \cite{douglas_wang_2013} proposed a modification to the search algorithm, where they used the shift operator for the UCDTQW on the $2^{n}$-complete graph with self-loops, $\mathcal{K}_{2^n}$, which consists of $n$ \textit{SWAP} gates. Furthermore, Wing and Venegas-Andraca \cite{wbva202301} developed a new shift operator for the UCDTQW on the graph $\mathcal{K}_{2^n}$ which consists of $n$ \textit{CNOT} gates, and \textit{CNOT} gates are native gates in IBM's NISQ computers. Thus, following the same logic as in \cite{douglas_wang_2013}, we will analyze the behavior of the search algorithm using as shift operator the model for the UCDTQW on $\mathcal{K}_{2^n}$ that was proposed in \cite{wbva202301}. A low-dimensional instance of both circuits is presented in Fig. \ref{model_circuit_example_both_cases}. In \cite{wbva202301} the model shown in fig \ref{model_circuit_example} was called the \textit{SWAP} model and the model shown in \ref{model_circuit_corollary_example} was called the \textit{CNOT} model.

To improve even further the experimental performance of the algorithm, we decided to use the Hadamard coin, given that the Hadamard coin is far less expensive than the Grover coin. The reason being is that the Hadamard coin applies a single-qubit Hadamard gate to each of the $m$ qubits of the coin register, while the quantum circuit implementation of Grover's operator uses an ($m-1$)-qubit multi-control Z gate, $C^{m-1}(Z)$, and the gate $C^{m-1}(Z)$ is typically decomposed into an exponential number of \textit{CNOT} and phase shift gates \cite{acasiete_agostini_moqadam_portugal_2020}, where $m$ is the size of the coin register. Therefore, as an example, we display in Fig. \ref{general_circuit_search_complement} the circuit form of the modified evolution operator $U'$ we will use for the search of state $\ket{1}^{\otimes n}$, which is associated to node $2^n-1$. To modify the target state of the search complement circuit we must modify the sequence of white and black controls of the perturbed coin operator. As explained in \cite{wbva202301}, a white control is a symbolic representation of a negated negated black control, and to negate a black control we add \textit{CNOT} gates to both sides of the black control. The sequence of white and black controls then can be associated to a binary string whose representation as an integer corresponds to the target node of the algorithm on the graph $\mathcal{K}_{2^n}$.

According to Algorithm \ref{quantum_walk_search_algorithm}, to execute the search algorithm, we have to set the initial state of the qubits to $\ket{0}$, apply single qubit Hadamard gates to all qubits of both coin and position registers, and to apply the circuit in Fig. \ref{general_circuit_search_complement} $\frac{\pi}{2}\sqrt{2^n}$ times. 

In the case when we perform one single step of the UCDTQW, the choice of the perturbed Hadamard coin in combination with the \textit{CNOT} model for the shift operator of the system leads to the opposite behavior of a search algorithm, i.e., the probability associated to the target node will be decreased. Thus the evolution operator we just created can be used as an evolution operator for the \textit{search complement} on $\mathcal{K}_{2^n}$. In the next subsection, we will study the search complement algorithm from the framework of directed multigraphs to provide an explanation for this behavior.

\section{\label{section_explanation_alg}Explanation of the Search Complement Algorithm from a Matrix and Multigraph Perspective}

In this section, we provide an explanation of the search complement algorithm from the point of view of multigraphs, using the matrix approach presented in section \ref{DTQW_multigraphs}. This approach allows us to study the dynamics of a walker on the graph associated to the perturbed evolution operator and provide an explanation for the algorithm.

\subsection{Search Complement Algorithm}

\begin{figure}[t!]
    \centering
\resizebox{\textwidth}{!}{    
    \subfigure[]{
    \centering
    \includegraphics[scale=0.65]{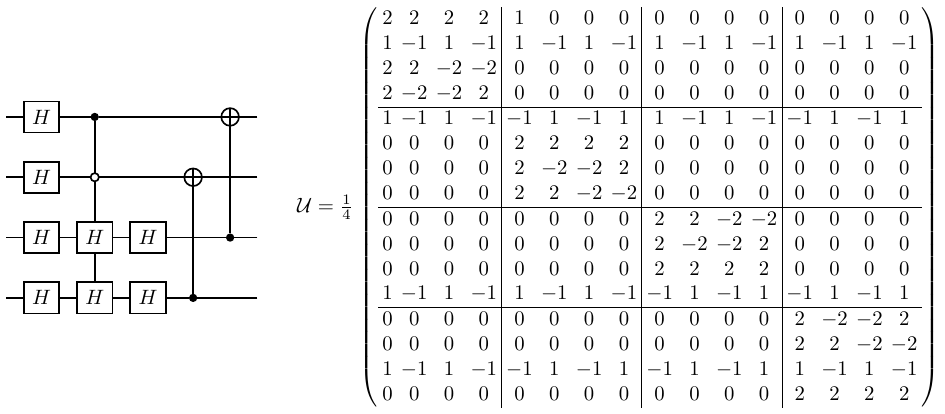}
    \label{search_hadamard_cnot_a}
    }
    \subfigure[]{
    \centering
    \includegraphics[scale=0.42]{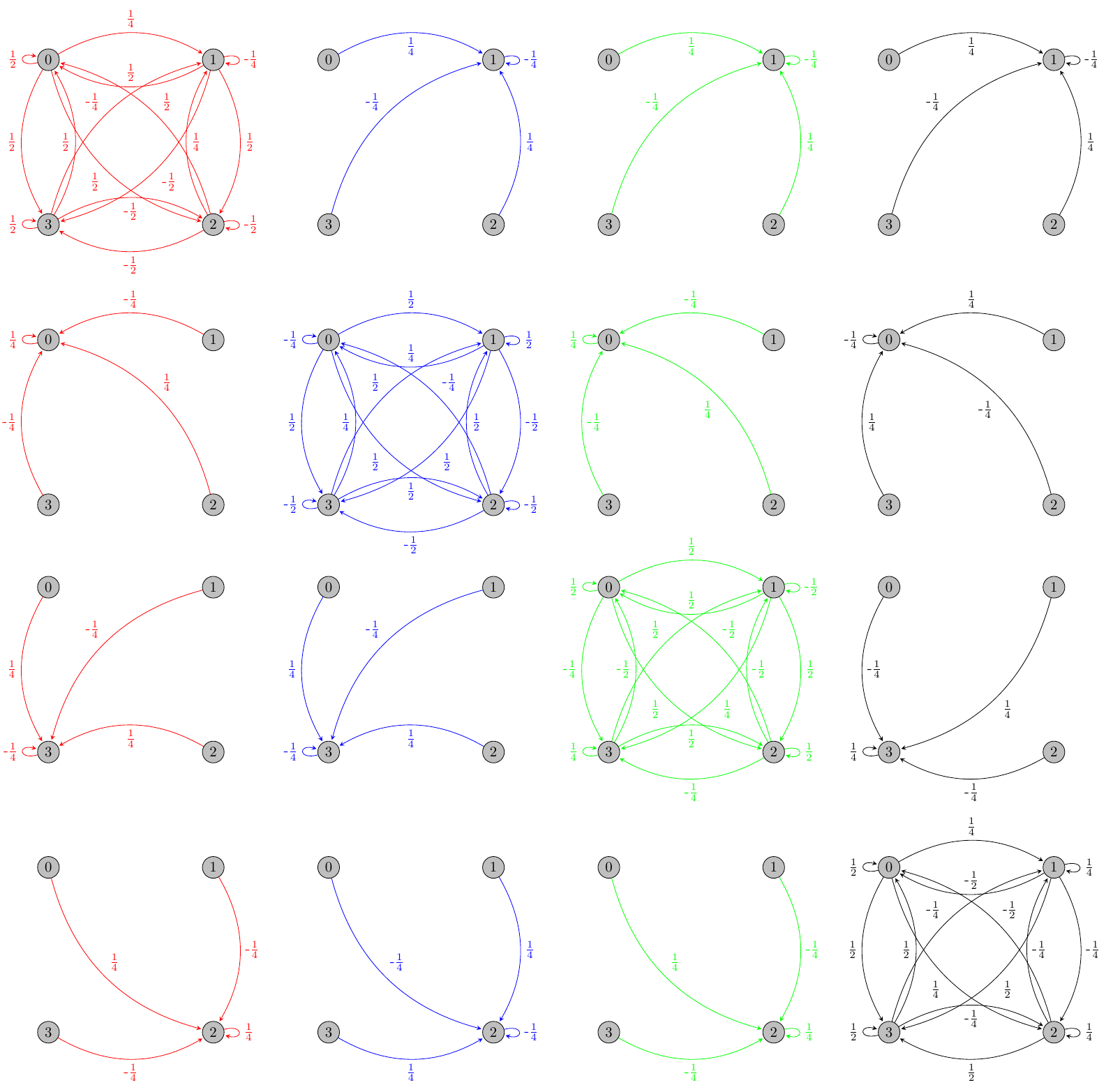}
    \label{search_hadamard_cnot_b}
    }
    }
    \caption{(a) Quantum circuit and matrix form of the evolution operator to perform the search complement algorithm of node 1 on $\mathcal{K}_4$. The pattern of black and white controls can be changed to search for other nodes. (b) Multigraph associated to quantum circuit in (a).}
    \label{search_hadamard_cnot}
\end{figure}

Consider the perturbed evolution operator, $U'$, for the search complement of node 1 on $\mathcal{K}_4$, which will serve us as an introductory example to explain the functioning of the algorithm and lead to generalizations. Given that to initialize the algorithm, we must apply a single-qubit Hadamard gate to all $2n$ qubits in both coin and position registers, we will analyze $\mathcal{U}=U'H^{\otimes (2n)}$, rather than $U'$ alone. Fig. \ref{search_hadamard_cnot} shows $\mathcal{U}$ in quantum circuit, block matrix and multigraph forms.

Notice that for visualization purposes, now we present separately the subgraph associated to each of the block operators in $\mathcal{U}$, rather than a single graph as in the example given in Fig. \ref{model_circuit_corollary_example}, although the idea is the same, i.e., we consider the union of the arcs associated to each block operator using the same set of vertices to obtain the multigraph associated to the operator $\mathcal{U}$, and the walker with coin state $|c_i\rangle$ moves along arcs corresponding to the block operators of the $i$th block column of $\mathcal{U}$. In Fig. \ref{search_hadamard_cnot_b}, the subgraphs associated to specific coin have a specific color. In this way, walkers with coin states $|00\rangle$, $|01\rangle$, $|10\rangle$ and $|11\rangle$ can move through red, blue, green and black arcs, respectively.

If we take a closer look at the red subgraphs in Fig. \ref{search_hadamard_cnot_b}, we can see that the number of incoming arcs is the same for all nodes except for node 1, which is the node associated to the target state in Fig. \ref{search_hadamard_cnot_b}. Only in the top red subgraph there  are arcs pointing towards node 1, and the weights of the arcs that point towards node 1 in the top subgraph are the ones with lowest weight in terms of absolute value. Furthermore, the incoming arcs to the nodes 0, 2 and 3 in the whole red subgraph come evenly from all nodes in the graph. To study the implications of this graph structure, we first choose to follow a matrix approach. This means that, if the walker starts the quantum walk on the red subgraphs, after the first step, the probability to find the walker at nodes 0, 2 or 3, will be evenly distributed, while the probability of finding the walker at node 1 will differ. To study in more detail the behavior of this graph structure we will choose a matrix approach.

Consider the action of $\mathcal{U}$ on $|\psi_0\rangle=|00\rangle_C\otimes|00\rangle_P$ as presented in Eq. \eqref{search_algorithm_action}.
\begin{equation}
\resizebox{0.9\textwidth}{!}{  
    $\mathcal{U}|\psi_0\rangle=\frac{1}{4}$
    $\left(
    \arraycolsep=2pt\def\arraystretch{1}
    \begin{array}{ *{4}{c} | *{4}{c} | *{4}{c} | *{4}{c} }
    2 & 2 & 2 & 2 & 1 & 0 & 0 & 0 & 0 & 0 & 0 & 0 & 0 & 0 & 0 & 0 \\
    1 & -1 & 1 & -1 & 1 & -1 & 1 & -1 & 1 & -1 & 1 & -1 & 1 & -1 & 1 & -1 \\ 
    2 & 2 & -2 & -2 & 0 & 0 & 0 & 0 & 0 & 0 & 0 & 0 & 0 & 0 & 0 & 0 \\
    2 & -2 & -2 & 2 & 0 & 0 & 0 & 0 & 0 & 0 & 0 & 0 & 0 & 0 & 0 & 0 \\ \hline
    1 & -1 & 1 & -1 & -1 & 1 & -1 & 1 & 1 & -1 & 1 & -1 & -1 & 1 & -1 & 1 \\ 
    0 & 0 & 0 & 0 & 2 & 2 & 2 & 2 & 0 & 0 & 0 & 0 & 0 & 0 & 0 & 0 \\
    0 & 0 & 0 & 0 & 2 & -2 & -2 & 2 & 0 & 0 & 0 & 0 & 0 & 0 & 0 & 0 \\ 
    0 & 0 & 0 & 0 & 2 & 2 & -2 & -2 & 0 & 0 & 0 & 0 & 0 & 0 & 0 & 0 \\ \hline
    0 & 0 & 0 & 0 & 0 & 0 & 0 & 0 & 2 & 2 & -2 & -2 & 0 & 0 & 0 & 0 \\        
    0 & 0 & 0 & 0 & 0 & 0 & 0 & 0 & 2 & -2 & -2 & 2 & 0 & 0 & 0 & 0 \\ 
    0 & 0 & 0 & 0 & 0 & 0 & 0 & 0 & 2 & 2 & 2 & 2 & 0 & 0 & 0 & 0 \\
    1 & -1 & 1 & -1 & 1 & -1 & 1 & -1 & -1 & 1 & -1 & 1 & -1 & 1 & -1 & 1 \\ \hline
    0 & 0 & 0 & 0 & 0 & 0 & 0 & 0 & 0 & 0 & 0 & 0 & 2 & -2 & -2 & 2 \\
    0 & 0 & 0 & 0 & 0 & 0 & 0 & 0 & 0 & 0 & 0 & 0 & 2 & 2 & -2 & -2 \\ 
    1 & -1 & 1 & -1 & -1 & 1 & -1 & 1 & -1 & 1 & -1 & 1 & 1 & -1 & 1 & -1 \\ 
    0 & 0 & 0 & 0 & 0 & 0 & 0 & 0 & 0 & 0 & 0 & 0 & 2 & 2 & 2 & 2 \\
   \end{array}
   \right)$
   
   $\left(
    \arraycolsep=2pt\def\arraystretch{1}
    \begin{array}{ *{1}{c}}
    1 \\
    0 \\
    0 \\
    0 \\ \hline
    0 \\
    0 \\
    0 \\
    0 \\ \hline
    0 \\
    0 \\
    0 \\
    0 \\ \hline
    0 \\
    0 \\
    0 \\ 
    0 \\
   \end{array}
   \right)=$
   
   $\left(
    \arraycolsep=2pt\def\arraystretch{1}
    \begin{array}{ *{1}{c}}
    1/2 \\
    1/4 \\
    1/2 \\
    1/2 \\ \hline
    1/4 \\
    0 \\
    0 \\
    0 \\ \hline
    0 \\
    0 \\
    0 \\
    1/4 \\ \hline
    0 \\
    0 \\
    1/4 \\ 
    0 \\
   \end{array}
   \right)$
}
\label{search_algorithm_action}
\end{equation}

In Eq. \eqref{search_algorithm_action} both the operator $\mathcal{U}$ and the state vector $|\psi_0\rangle$ are split into block matrices and subvectors, respectively, according to the coin state of the quantum walk. The $j$th entry of the $i$th subvector corresponds to the state $|c_i\rangle\otimes|{v_j}\rangle$, thus the vector associated to $|\psi_0\rangle=|00\rangle\otimes|00\rangle$ has a 1 in the zeroth entry of the zeroth subvector, and zeros in the rest of the entries. Therefore, the resulting state of the operation $\mathcal{U}|\psi_0\rangle$ is an extraction of the zeroth column of the operator $\mathcal{U}$. Following Eq. \eqref{probability_vector_hadamard} on $\mathcal{U}|\psi_0\rangle$, we get the probability distribution vector is given by
\begin{equation}
P=
\begin{pmatrix}
\frac{1}{2}^{*}\cdot\frac{1}{2}+\frac{1}{4}^{*}\cdot\frac{1}{4}\\
\frac{1}{4}^{*}\cdot\frac{1}{4}\\
\frac{1}{2}^{*}\cdot\frac{1}{2}+\frac{1}{4}^{*}\cdot\frac{1}{4}\\
\frac{1}{2}^{*}\cdot\frac{1}{2}+\frac{1}{4}^{*}\cdot\frac{1}{4}\\
\end{pmatrix}=\begin{pmatrix}
5/16\\
1/16\\
5/16\\
5/16\\
\end{pmatrix}
\label{probability_distribution_1st_step_k_4}
\end{equation}
From Eq. \eqref{probability_distribution_1st_step_k_4} we can see that the probability of measuring the position state $|01\rangle$, which is associated to node 1, is indeed reduced. 

So far in our example, we have shown that the algorithm works for the case in which we initialize the walker with the quantum state $|\psi\rangle=|00\rangle_C\otimes|00\rangle_P$. Let us now analyze the behavior of the algorithm under different initial states. For this purpose, we will calculate the probability matrix associated to $\mathcal{U}$, which contains the probability vectors associated to all combinations of initial states in the UCDTQW.

First, we calculate $\mathcal{U}^{*} \odot \mathcal{U}$ in Eq. \eqref{hadamard_product_U_cnot}. $\mathcal{U}^{*} \odot \mathcal{U}$ transforms the probability amplitudes associated to $\mathcal{U}$ to probabilities.

\begin{equation}
\resizebox{0.8\textwidth}{!}{  
$   \mathcal{U}^{*}\odot\mathcal{U}=
    \left(
    \arraycolsep=2pt\def\arraystretch{1}
    \begin{array}{ *{4}{c} | *{4}{c} | *{4}{c} | *{4}{c} }
    1/4 & 1/4 & 1/4 & 1/4 & 0 & 0 & 0 & 0 & 0 & 0 & 0 & 0 & 0 & 0 & 0 & 0 \\
    1/16 & 1/16 & 1/16 & 1/16 & 1/16 & 1/16 & 1/16 & 1/16 & 1/16 & 1/16 & 1/16 & 1/16 & 1/16 & 1/16 & 1/16 & 1/16 \\ 
    1/4 & 1/4 & 1/4 & 1/4 & 0 & 0 & 0 & 0 & 0 & 0 & 0 & 0 & 0 & 0 & 0 & 0 \\
    1/4 & 1/4 & 1/4 & 1/4 & 0 & 0 & 0 & 0 & 0 & 0 & 0 & 0 & 0 & 0 & 0 & 0 \\ \hline
    1/16 & 1/16 & 1/16 & 1/16 & 1/16 & 1/16 & 1/16 & 1/16 & 1/16 & 1/16 & 1/16 & 1/16 & 1/16 & 1/16 & 1/16 & 1/16 \\
    0 & 0 & 0 & 0 & 1/4 & 1/4 & 1/4 & 1/4 & 0 & 0 & 0 & 0 & 0 & 0 & 0 & 0 \\
    0 & 0 & 0 & 0 & 1/4 & 1/4 & 1/4 & 1/4 & 0 & 0 & 0 & 0 & 0 & 0 & 0 & 0 \\
    0 & 0 & 0 & 0 & 1/4 & 1/4 & 1/4 & 1/4 & 0 & 0 & 0 & 0 & 0 & 0 & 0 & 0 \\ \hline
    0 & 0 & 0 & 0 & 0 & 0 & 0 & 0 & 1/4 & 1/4 & 1/4 & 1/4 & 0 & 0 & 0 & 0 \\        
    0 & 0 & 0 & 0 & 0 & 0 & 0 & 0 & 1/4 & 1/4 & 1/4 & 1/4 & 0 & 0 & 0 & 0 \\ 
    0 & 0 & 0 & 0 & 0 & 0 & 0 & 0 & 1/4 & 1/4 & 1/4 & 1/4 & 0 & 0 & 0 & 0 \\
    1/16 & 1/16 & 1/16 & 1/16 & 1/16 & 1/16 & 1/16 & 1/16 & 1/16 & 1/16 & 1/16 & 1/16 & 1/16 & 1/16 & 1/16 & 1/16 \\ \hline
    0 & 0 & 0 & 0 & 0 & 0 & 0 & 0 & 0 & 0 & 0 & 0 & 1/4 & 1/4 & 1/4 & 1/4 \\
    0 & 0 & 0 & 0 & 0 & 0 & 0 & 0 & 0 & 0 & 0 & 0 & 1/4 & 1/4 & 1/4 & 1/4 \\
    1/16 & 1/16 & 1/16 & 1/16 & 1/16 & 1/16 & 1/16 & 1/16 & 1/16 & 1/16 & 1/16 & 1/16 & 1/16 & 1/16 & 1/16 & 1/16 \\
    0 & 0 & 0 & 0 & 0 & 0 & 0 & 0 & 0 & 0 & 0 & 0 & 1/4 & 1/4 & 1/4 & 1/4 \\
   \end{array}
   \right)$
}   
\label{hadamard_product_U_cnot}
\end{equation}

Next, we add all the block rows in $\mathcal{U}^{*} \odot \mathcal{U}$, to obtain the probability matrix associated to $\mathcal{U}$, as shown in Eq. \eqref{probability_matrix_search_complement_k4}. 

\begin{equation}
\resizebox{0.8\textwidth}{!}{  
$   M_P=
    \left(
    \arraycolsep=2pt\def\arraystretch{1}
    \begin{array}{ *{4}{c} | *{4}{c} | *{4}{c} | *{4}{c} }
    5/16 & 5/16 & 5/16 & 5/16 & 1/16 & 1/16 & 1/16 & 1/16 & 5/16 & 5/16 & 5/16 & 5/16 & 5/16 & 5/16 & 5/16 & 5/16 \\
    1/16 & 1/16 & 1/16 & 1/16 & 5/16 & 5/16 & 5/16 & 5/16 & 5/16 & 5/16 & 5/16 & 5/16 & 5/16 & 5/16 & 5/16 & 5/16 \\
    5/16 & 5/16 & 5/16 & 5/16 & 5/16 & 5/16 & 5/16 & 5/16 & 5/16 & 5/16 & 5/16 & 5/16 & 1/16 & 1/16 & 1/16 & 1/16 \\
    5/16 & 5/16 & 5/16 & 5/16 & 5/16 & 5/16 & 5/16 & 5/16 & 1/16 & 1/16 & 1/16 & 1/16 & 5/16 & 5/16 & 5/16 & 5/16 \\ 
   \end{array}
   \right)$
}   
   \label{probability_matrix_search_complement_k4}
\end{equation}

\begin{figure}[h!]
    \centering
\resizebox{0.8\textwidth}{!}{      
    \includegraphics[scale=0.8]{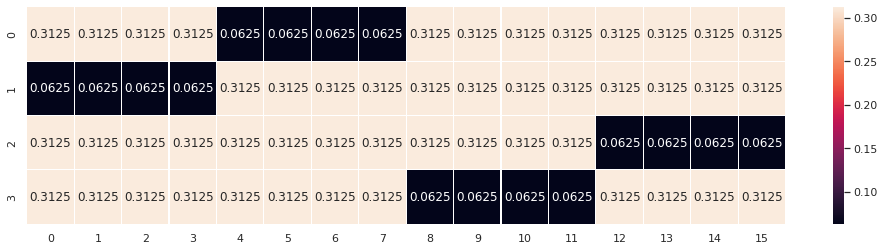}
}
    \caption{Probability matrix, $M_P$, of the complement of a search on the graph $\mathcal{K}_4$. The columns of the probability matrix contain the probability distribution associated to all the possible combinations of initial coin and position states of a quantum walk. It is calculated by performing the product $\mathcal{U}^{*}\odot\mathcal{U}$, and then adding the block rows of $\mathcal{U}^{*}\odot\mathcal{U}$, such that all block rows have dimension $n \times nm$, where $n$ and $m$ are the dimensions of the position and coin registers, respectively. The indices in the columns of $M_P$ correspond to the decimal representation of the initial state of the walker $|\psi_0\rangle$.}
    \label{probability_matrix_complement_target1}
\end{figure}

Using Qiskit \cite{Qiskit} we can generate the unitary matrix form of a circuit and then perform the necessary operations to collapse it into a probability matrix and visualize the probability matrix as a heatmap for better visual appreciation. Fig. \ref{probability_matrix_complement_target1} displays the probability matrix associated to operator $\mathcal{U}$ in Eq. \eqref{search_algorithm_action}.

\begin{figure}[h!]
    \centering
\resizebox{0.85\textwidth}{!}{    
\includegraphics[]{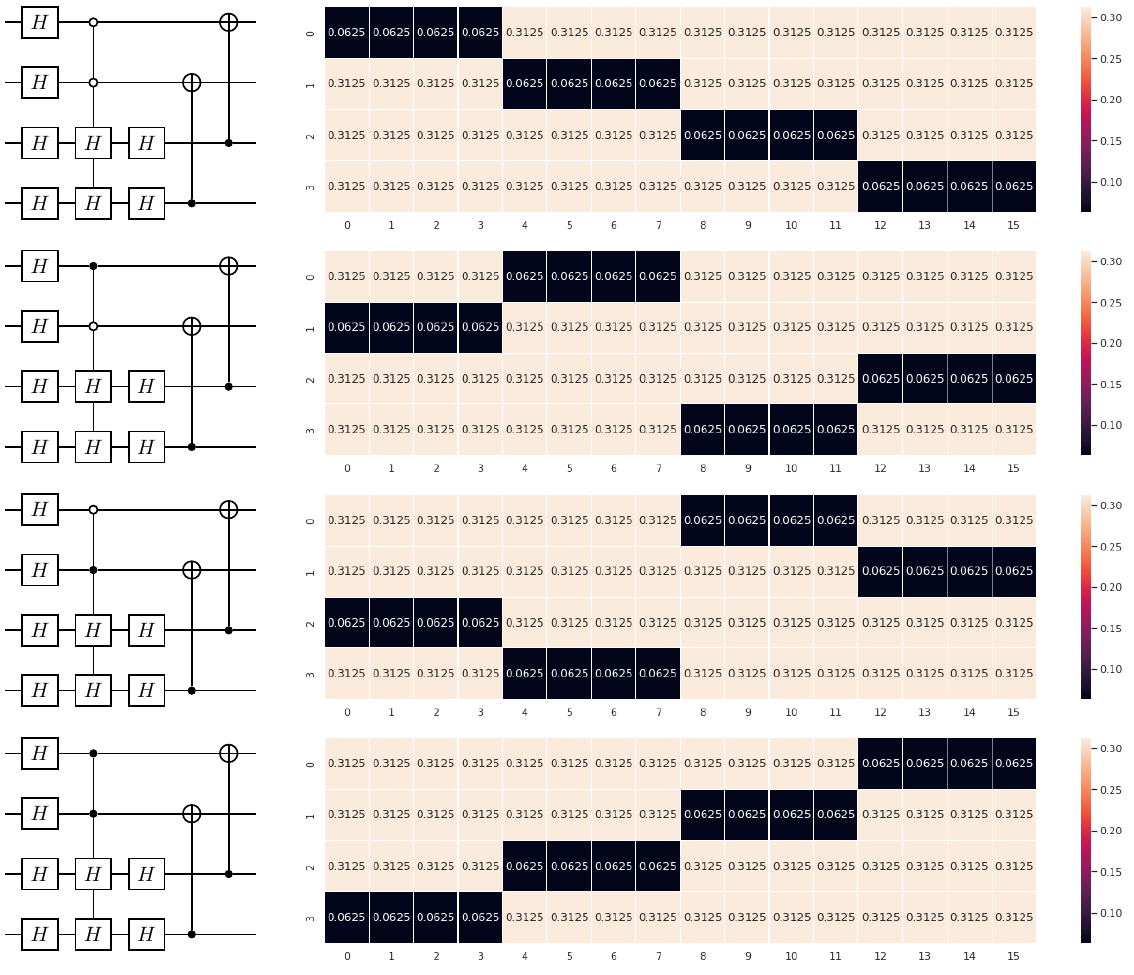}
}
    \caption{Circuit implementation of the search complement on $\mathcal{K}_4$ for different target nodes along side with their corresponding probability matrix. The target node of the circuits is 1, 2, 3 and 4 from top to bottom.}
    \label{search_complement_probability_matrix}
\end{figure}

Now, the calculation of the probability matrix might be useful to study the behaviour of the algorithm in low-dimensional cases of different topologies, using different coins, different target nodes and different initial states, in order for us to notice patterns that can generalize to higher-dimensional cases. As an example of this statement, consider Fig. \ref{search_complement_probability_matrix}. This figure contains the search complement circuit for target states $\ket{00}$, $\ket{01}$, $\ket{10}$ and $\ket{11}$, in that order from top to bottom, and next to each circuit we can see the associated probability matrix. The probability matrices are split into blocks delimited by four consecutive columns. The blocks of the matrices are associated to  coin states $\ket{00}$, $\ket{01}$, $\ket{10}$ and $\ket{11}$, in that order from left to right, and  each column within a block is associated to position states $\ket{00}$, $\ket{01}$, $\ket{10}$ and $\ket{11}$, in that order from left to right too.

The search complement circuits in Fig. \ref{search_complement_probability_matrix} have as target the states $|00\rangle$, $|01\rangle$, $|10\rangle$ and $|11\rangle$, respectively from top to bottom. The only columns that have a low probability in the entry that corresponds to the target state, are the first four columns of each probability matrix. The first four columns of each probability matrix are associated to coin state $|00\rangle$. This means that for any of the quantum circuits of Fig. \ref{search_complement_probability_matrix} if we initialize the state the UCDTQW $\ket{\psi}$ with any position state, but always keep the coin state fixed at $|00\rangle$, i.e. $|\psi_0\rangle=|00\rangle_C|q_1q_0\rangle_P$, the operation between the basis vector state associated to $\ket{\psi}$ and the probability matrix $M_P$ associated to the quantum circuit, will be to extract one of the first four columns of $M_P$, and thus extract one of the probability vectors that reduce the probability of obtaining the target state. That is to say, if $\mathcal{U}$ is the perturbed evolution operator that minimizes the probability of measuring node $j$, for simplicity, we can always set $|\psi_0\rangle=|00\rangle_C|00\rangle_P$ to get the desired result.

\begin{figure}[h!]
    \centering
\resizebox{\textwidth}{!}{      
    \includegraphics[scale=1]{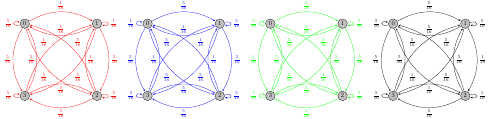}
}
    \caption{Collapse of the multigraph associated to the search complement circuit shown that has node 1 as target (see Fig. \ref{search_hadamard_cnot_b})}
    \label{collapse_of_multigraph_k4}
\end{figure}

As described in section \ref{DTQW_multigraphs}, if $\mathcal{U}$ is a bipartite operator associated to a multigraph $\mathcal{G}$, then the probability matrix associated to $\mathcal{U}$ corresponds to a \textit{collapsed} version of $\mathcal{G}$, which we will refer to as $\mathcal{G}_c$. $\mathcal{G}_c$ is a multigraph such that it condenses all the arcs associated to coin state 
$\ket{k}$ that go out from vertex $i$ and come into vertex $j$ of $\mathcal{G}$ into a single one whose weight is the sum of the square modulus the arcs in $\mathcal{G}$ that go out from vertex $i$ and come into vertex $j$. For example, the collapsed version of the multigraph associated to the search complement circuit with target state $\ket{01}$ (Fig. \ref{search_hadamard_cnot_b}) is presented in Fig. \ref{collapse_of_multigraph_k4}.

Fig. \ref{collapse_of_multigraph_k4} is a multigraph that contains four complete graphs as subgraphs, each one associated to a coin state of the system.  From this graph we can see that the arcs that come into node 1 in the red subgraph are the ones with lower weight. Thus if we perform one step and then measure, the probability of finding the walker in node 1 will be the lowest with respect to the other nodes. If we were to trace the collapsed multigraphs associated to the probability matrix corresponding to the complement search circuit of nodes 0, 2 and 3 (shown in Fig. \ref{search_complement_probability_matrix}), then we would see the same pattern, i.e. for the red subgraphs, the node with lowest in-degree would be node 0, 2 and 3, respectively.

Thus, in simple terms, the core functioning of our algorithm is creating a matrix associated to a graph in which the in- and out-degrees of all nodes are the evenly distributed, except for the in-degree of the target node, which must be lower than the rest, and the apply it to a basis vector.

In Appendix \ref{Formal Proof of the Search Complement Algorithm} we provide a general proof that the behavior of the quantum walk continues for larger-dimensional $\mathcal{K}_{2^n}$ graphs. The implication is that as we increase the number of nodes, we increase the number of non-target states and also the number of arcs with larger weights pointing towards non-target nodes. Thus, after performing one step of the UCDTQW and measure, the quantum walker will have a larger probability to be found in any non-target node. Therefore, the idea of this algorithm is to keep increasing the size of the graph, until we approximate the target node's probability to zero.

\begin{figure}[b!]
    \centering
\resizebox{1\textwidth}{!}{    
\includegraphics[]{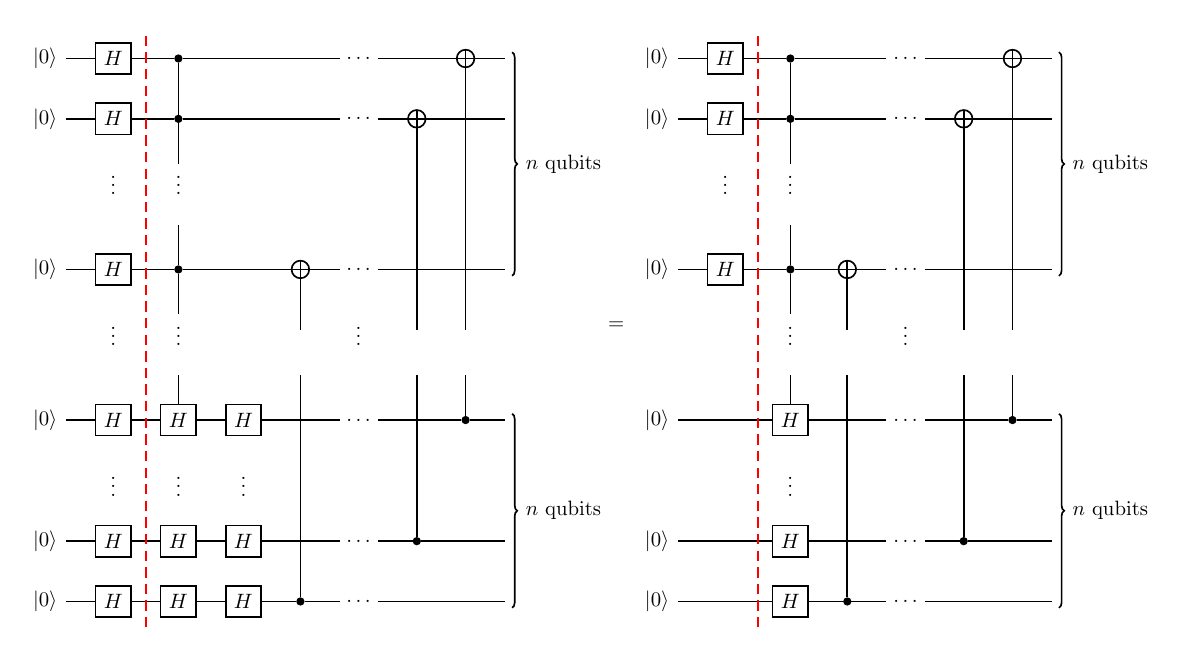}
}
    \caption{Synthesis of the quantum circuit associated to the search complement of node $2^n-1$ on the graph $\mathcal{K}_{2^n}$. From the synthesized circuit at the right-hand side of the figure, we can see that only the position register needs to be put into a superposition of states for the algorithm to work. From multigraph theoretical framework, that means the quantum walk takes place in the subgraph associated to coin state $\ket{0}^{\otimes n}$.}
    \label{general_search_complement_circuit}
\end{figure}

Next we preset an explanation of the search complement algorithm as an effect of quantum superposition from a quantum circuit approach.

In Fig. \ref{general_search_complement_circuit} we show an example of the general circuit for search complement algorithm that where the 
$C^n(H^{\otimes n})$ gate activates when the position register is $\ket{1}^{\otimes n}$, that is, $C^n(H^{\otimes n})$ acts as an oracle that encodes the state associated to node $2^n-1$. Furthermore, in Fig. \ref{general_search_complement_circuit}, we present a synthesis of the circuit associated to the algorithm which simplifies the proof and can also help us achieve better experimental results.

Consider the general case when the controls of $C^n(H^{\otimes n})$ encode a state associated to node $t$. From now on we will use the notation introduced \cite{wbva202301}, where multi-control gates have two superindices. The first superindex corresponds to the number of controls and the second superindex corresponds to the decimal representation of the sequence formed by the controls of the quantum gate. Thus, the multi-control Hadamard gate in Fig. \ref{general_search_complement_circuit} can be expressed by $C^{n, t}(H^{\otimes n})$. The quantum circuit implementation of the search complement algorithm consists of five steps:

\begin{enumerate}

\item We initialize the state of the system as $\ket{\psi_0} = \ket{0}^{\otimes n}\otimes\ket{q_{2^n-1}\dots q_1q_0}$, where $q_i\in \{0,1\}$. For simplicity, let us set the coin state to $\ket{0}^{\otimes n}$. Also let us express the coin and position states in decimal notation. Then, $\ket{\psi_0} = \ket{0}\otimes\ket{0}$.

\item We apply a Hadamard operator to all the qubits of the position register causing an equal superposition of all $n$-qubit states in the position register, i.e. 

\begin{equation}
    \ket{\psi_0'} = I_m\otimes H^{\otimes n}\ket{\psi} = \ket{0}\otimes \left( \frac{1}{\sqrt{2^n}}\sum\limits_{j = 0}^{2^n-1}\ket{j}\right)
\end{equation} 

\item We apply the gate $C^{n, t}(H^{\otimes n})$ to $\ket{\psi_0'}$, putting the coin register in a superposition of all states only when the position state coincides with state $\ket{t}$ encoded in the gate $C^{n, t}(H^{\otimes n})$. Now both coin and position registers are in a superposition of states, i.e.

\begin{align}
\ket{\psi''_0} = C^{n, t}{\bf (H^{\otimes n}})\ket{\psi'_0} \nonumber \\ 
 & \hspace{-2.58cm} = \ \ket{0} \otimes \left( \sum\limits_{\substack{j = 0, \\ j \neq t}}^{2^n-1} \frac{1}{\sqrt{2^n}} \ket{j}\right) + \left( \sum\limits_{i = 0}^{2^n-1} \frac{1}{\sqrt{4^n}}\ket{i}\right) \otimes \ket{t}
\label{algorithm_third_step}
\end{align}

\item We apply the sequence of $CNOT$ gates in Fig. \ref{general_search_complement_circuit} on $\ket{\psi_0''}$. The mathematical form of this sequence of gates was introduced in \cite{wbva202301}, and is shown in Eq. \eqref{shift_cnot_model}.

\begin{equation}
    S = \sum_{i=0}^{2^m-1} \sum_{j=0}^{2^n-1} \ket{i}\bra{i}\otimes \ket{j\oplus i}\bra{j}
\label{shift_cnot_model}
\end{equation}

The application of the sequence of $CNOT$ gates in Fig. \ref{general_search_complement_circuit}, or $S$, on $\ket{\psi''_0}$ only modifies the second term, and the effect it has is to flip the i$th$ qubit of the position state when the i$th$ qubit of the coin register is $\ket{1}$. In explicit binary notation, the position state in all the composite states of the second term of Eq. \eqref{algorithm_third_step} become $\ket{t_0t_1\dots t_{2^{n-1}} \oplus c_0c_1\dots c_{2^{n-1}}}$, where $\ket{c_i}$ is the i$th$ qubit of the coin register. As a consequence the second term of Eq. \eqref{algorithm_third_step} is transformed into the second term of Eq. \eqref{algorithm_fourth_step}.

\begin{equation}
\ket{\psi_1} = \ket{0} \otimes \left( \sum\limits_{\substack{j = 0, \\ j \neq t}}^{2^n-1} \frac{1}{\sqrt{2^n}} \ket{j}\right) + \sum\limits_{i = 0}^{2^n-1} \frac{1}{\sqrt{4^n}} \ket{i} \otimes \ket{t \oplus i} 
\label{algorithm_fourth_step}
\end{equation}

Step four in the application of the evolution operator is crucial given that the second term in Eq. \eqref{algorithm_third_step} is transformed into a superposition of composite states where each composite state takes different values for the position state. As a consequence, the probability amplitudes of the composite states in the second term of Eq. \eqref{algorithm_third_step} that were once associated to the target state only, are now redistributed on the second term of Eq. \eqref{algorithm_fourth_step} to composite states where the position state is not only the target. In fact, there is only one composite state where the position state is $\ket{t}$ in Eq. \eqref{algorithm_fourth_step}, and has probability amplitude $\frac{1}{\sqrt{4^n}}$.

\item We measure only the position register of system.
According to Eq. \eqref{node_probability_eq}, the probability of measuring node $k$ is given by $p(v_k) = \langle \psi(t)|M^{\dagger}_k M_k| \psi(t) \rangle$, where $M_k = I_m\otimes|k\rangle\langle k|$. Therefore, to calculate the probability of measuring the position states of the system, we first apply $M_k$ to $\ket{\psi_1}$, i.e.

\begin{equation}
    M_k\ket{\psi_1} = I_m\otimes\ket{k}\bra{k}\left[  \ket{0} \otimes \left( \sum\limits_{\substack{j = 0, \\ j \neq t}}^{2^n-1} \frac{1}{\sqrt{2^n}} \ket{j}\right) + \sum\limits_{i = 0}^{2^n-1} \frac{1}{\sqrt{4^n}} \ket{i} \otimes \ket{t \oplus i}\right] 
    \label{measurement_qc_proof}
\end{equation}

Then, we have two cases:
\begin{enumerate}
    \item When $\ket{k} \neq \ket{t}$
    \begin{equation}
        M_k\ket{\psi_1}M_k\ket{\psi_1} = \frac{1}{\sqrt{2^n}}\ket{0} \otimes \ket{k} + \frac{1}{\sqrt{4^n}}\ket{i_k} \otimes \ket{k} 
    \end{equation}
    where $i_k$ is the index that complies with the relation $k = t\oplus i$.

    \item When $\ket{k} = \ket{t}$:
    \begin{equation}
        M_t\ket{\psi_1} = \frac{1}{\sqrt{4^n}}\ket{0} \otimes \ket{t}
    \end{equation}

\end{enumerate}

Then, the probability of measuring non-target states is $p_{k\neq t} = \bra{\psi_1}M_kM_k\ket{\psi_1} = \frac{1}{2^n} + \frac{1}{4^n}$ and the probability of measuring the target state is $p_{t} = \bra{\psi_1}M_tM_t\ket{\psi_1} = \frac{1}{4^n}$. Thus, in general, the probability distribution of measuring the state associated to node $k$ is given by 

\begin{align}
    p(v_k) = \begin{cases}
        \frac{1}{4^{n}} &\text{if $k=t\oplus r$}\\
        \frac{1}{4^n}+\frac{1}{2^n} &\text{if $k \neq t\oplus r$}
    \end{cases}
\label{general_prob_dist_complement}
\end{align}


\end{enumerate}

Different choices for the initial state of the system might lead to different interference effects that do not decrease the probability of the target state, however it suffices to initialize the whole quantum register with the state $\ket{\psi} = \ket{0} \otimes \ket{0}$ for the search algorithm to work. A general and formal proof of this algorithm is presented in Appendix \ref{Formal Proof of the Search Complement Algorithm}. In the general proof we calculate the probability distribution associated to any initial state of the system and any target state.

\section{\label{section_experimentation_alg}Execution of the Search Complement Algorithm}

In this section we will present the results of the execution of the search complement algorithms on Qiskit, using the QASM simulator, and on IBM Quantum Composer using the quantum processor $ibmq\_lima$. A repository with the code used to perform the simulations is provided in \cite{github_repo_search_complement}.

\begin{figure}[h!]
    \centering

\subfigure[]{
\resizebox{0.47\textwidth}{!}{ 
\includegraphics{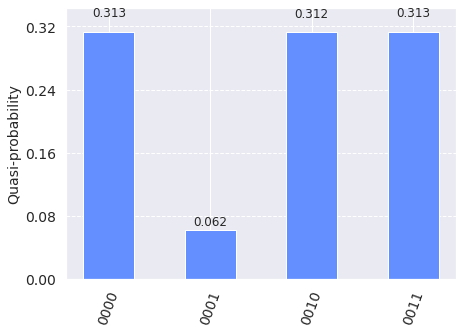}
}
\label{search_complement_236_a}
}
\subfigure[]{
\resizebox{0.47\textwidth}{!}{ 
\includegraphics{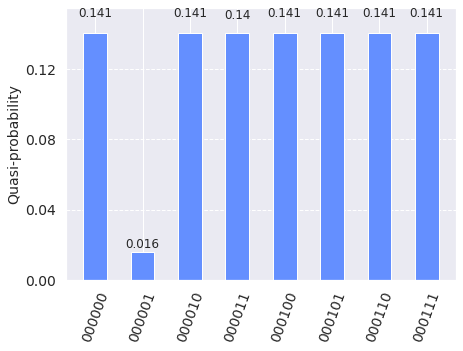}
}
\label{search_complement_236_b}
}

\subfigure[]{
\resizebox{\textwidth}{!}{ 
\includegraphics{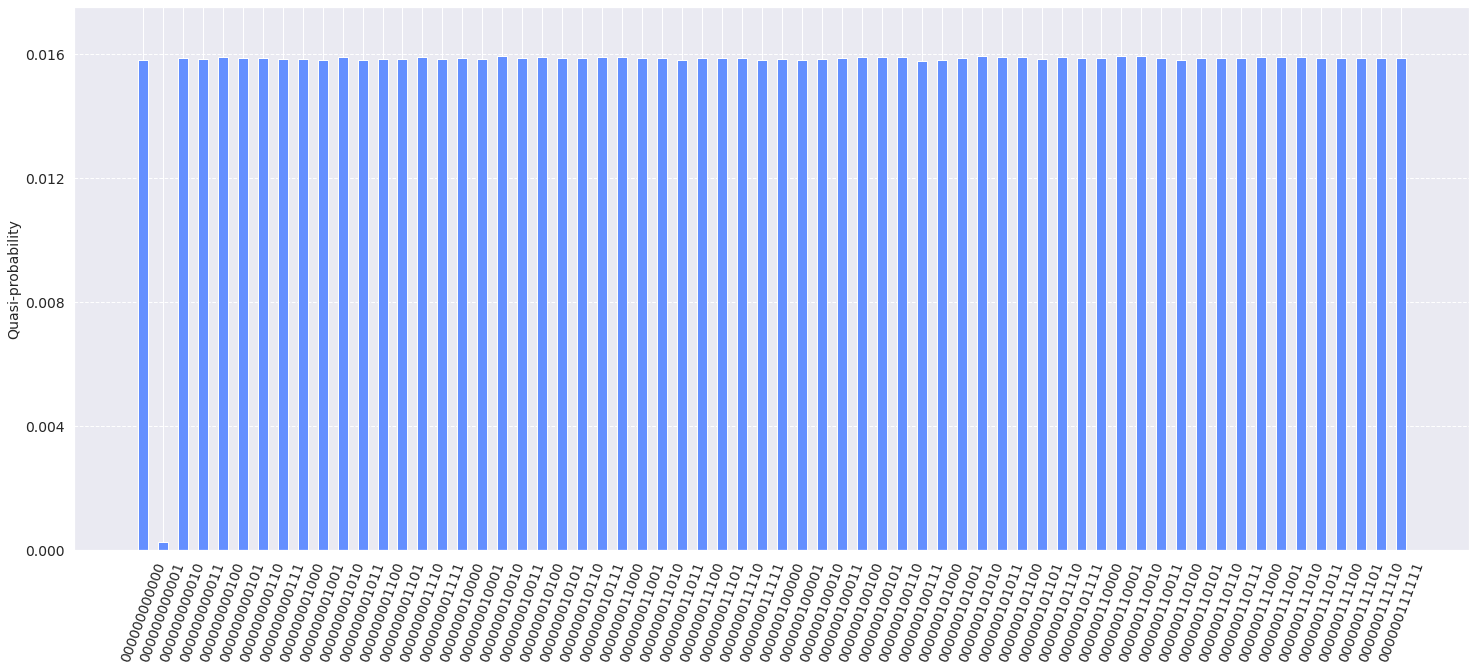}
}
\label{search_complement_236_c}
}
    \caption{Qiskit simulations of the search complement algorithm. We performed 1 step using a larger-scale version of the circuit proposed in Fig. \ref{search_hadamard_cnot}. In all the cases node 1 is the target. (a) Displays the probability distribution of the algorithm for a $\mathcal{K}_4$ graph, and using 8192 shots. In (b) we consider a $\mathcal{K}_8$ graph and one million shots and in (c) we consider a $\mathcal{K}_{64}$ graph and ten million shots.}
    \label{search_complement_236}
\end{figure}

In Fig. \ref{search_complement_236} we present the probability distribution of the walker after applying the search complement circuit associated to graphs $\mathcal{K}_4$, $\mathcal{K}_8$ and $\mathcal{K}_{64}$, which were simulated on Qiskit. In Fig. \ref{search_complement_236_a} the probability of measuring the target state is $6.2\%$, which is $\thicksim 5$ times smaller than the probability of measuring non-target states ($31.3\%$). In Fig. \ref{search_complement_236_b} the probability of measuring the target state is $1.6\%$, which is $\thicksim 8.8$ times smaller than the probability of measuring non-target states ($14.1\%$). Finally, in Fig. \ref{search_complement_236_c}, the probability of getting the state corresponding to node 1, becomes $0.02\%$, which is $\thicksim 80$ times smaller than then probability of measuring non-target states ($1.6\%$). The number of shots needed to obtain the probability distributions shown in Figures \ref{search_complement_236_a}, \ref{search_complement_236_b} and \ref{search_complement_236_c} were 8192, $10^6$ and $10^7$, respectively. 

The results shown in Fig. \ref{search_complement_236} provides experimental proof that the probability of the target state will decrease drastically faster than the probability of non-target states, however, the number of shots needed to obtain an experimental probability distribution that follows Eq. \eqref{general_prob_dist_complement} also increases drastically.

To end this section, we will present the experimental results of implementing the search complement algorithm on $\mathcal{K}_4$ with target node 3 on IBM's quantum processor $ibmq\_manila$ using the Quantum Composer.

\begin{figure}[h!]
    \centering
    \subfigure[]{
    \includegraphics[scale=0.8]{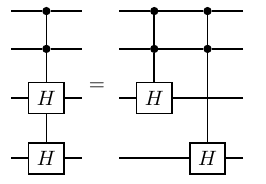}
    \label{decomposition_multicontrol_hadamard_a}
    }
    \subfigure[]{
    \includegraphics[scale=0.8]{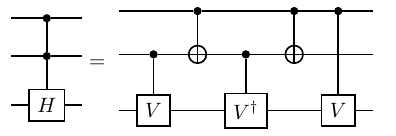}
    \label{decomposition_multicontrol_hadamard_b}
    }    
    \caption{(a) Decomposition of multi-control-multi-target Hadamard gate into two multi-control-single target gates. (b) Decomposition of multi-control-single target Hadamard gate, where $V=\sqrt{H}$. Combining both decompositions we can find an elementary gate representation of a multi-control-multi-target Hadamard gate.}
    \label{decomposition_multicontrol_hadamard}
\end{figure}

To begin with, currently, IBM's Quantum Composer does not allow the implementation of multi-target-multi-controlled Hadamard gates. Thus, we have to apply gate identities to decompose it into single-qubit and single-control single-target gates. The identities used are the one displayed in Fig. \ref{decomposition_multicontrol_hadamard}, where we first present a way to decompose the original gate into two multiple-control single-target gates (Fig. \ref{decomposition_multicontrol_hadamard_a}), and then we present how to decompose a two-control single-target gate into two CNOT gates and two single-control single-target gates with target $V=\sqrt{H}$ (Fig. \ref{decomposition_multicontrol_hadamard_b}). The quantum composer is able to implement a general single-control single-target gate of the form given in Eq. \eqref{general_controlled_gate}, if the angles $\theta, \phi$ and $\lambda$ are given.

\begin{equation}
C_U(\theta,\phi,\lambda)=
\begin{pmatrix}
1 & 0 & 0 & 0 \\
0 & e^{-i(\phi+\lambda)/2}cos(\theta/2) & 0 & e^{-i(\phi-\lambda)/2}sin(\theta/2) \\
0 & 0 & 1 & 0 \\
0 & e^{i(\phi-\lambda)/2}sin(\theta/2) & 0 & e^{i(\phi+\lambda)/2}cos(\theta/2) 
\end{pmatrix}
\label{general_controlled_gate}
\end{equation}

\begin{figure}[h!]
    \centering
\resizebox{0.9\textwidth}{!}{    
\includegraphics[]{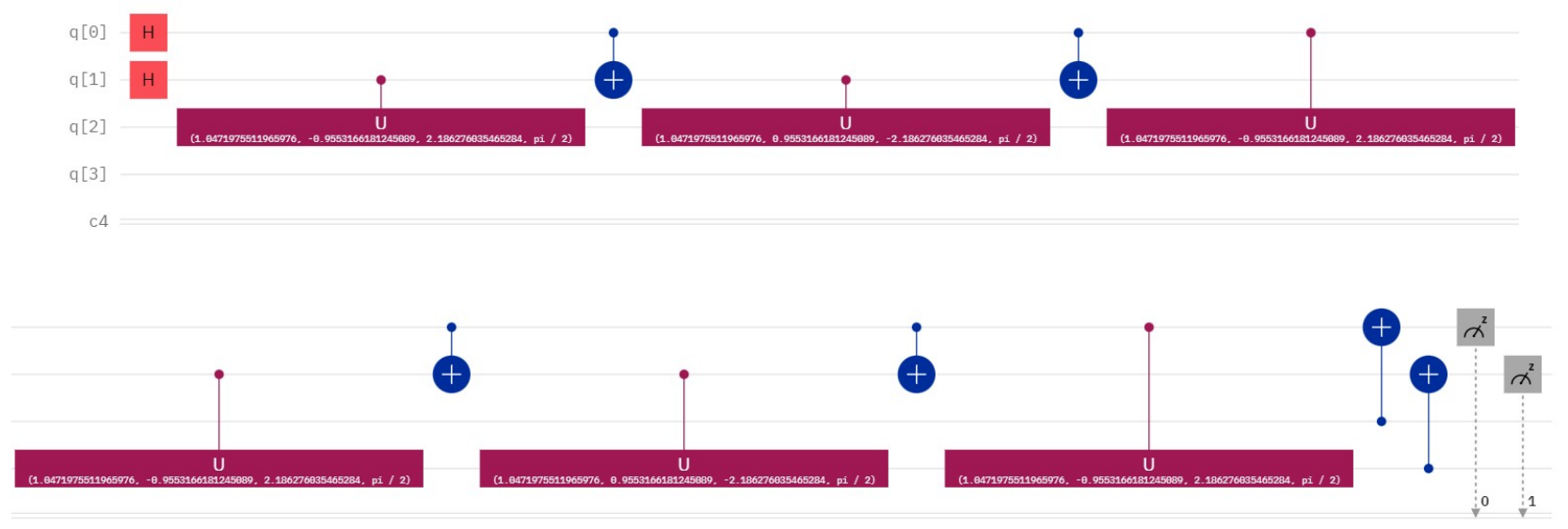}
}
    \caption{Quantum circuit implemented of the search complement algorithm on $\mathcal{K}_{4}$ with node 3 as target node on IBM Quantum Composer. To modify the target node of the circuit we must negate the controls of the decomposed Hadamard operator using \textit{CNOT} gates to both sides of the control we want to negate.}
    \label{search_k4_composer}
\end{figure}

Qiskit's \textit{quantum\_info} module contains the function \textit{OneQubitEulerDecomposer}, which returns the angles $\theta, \phi$ and $\lambda$ for an arbitrary unitary matrix. In this way, we can find that the angles are $\theta=1.0471975511965976,\; \phi=-0.9553166181245089$ and $\lambda=2.186276035465284$ for $\sqrt{H}$ and $\theta=1.0471975511965976,\; \phi=0.9553166181245089$ and $\lambda=-2.186276035465284$ for $(\sqrt{H})^{\dagger}$. Thus, the circuit implementation of the search complement algorithm on $\mathcal{K}_4$ with target node 3, is displayed in Fig. \ref{search_k4_composer}.

\begin{figure}[h!]
    \centering
\resizebox{0.9\textwidth}{!}{    
\includegraphics[]{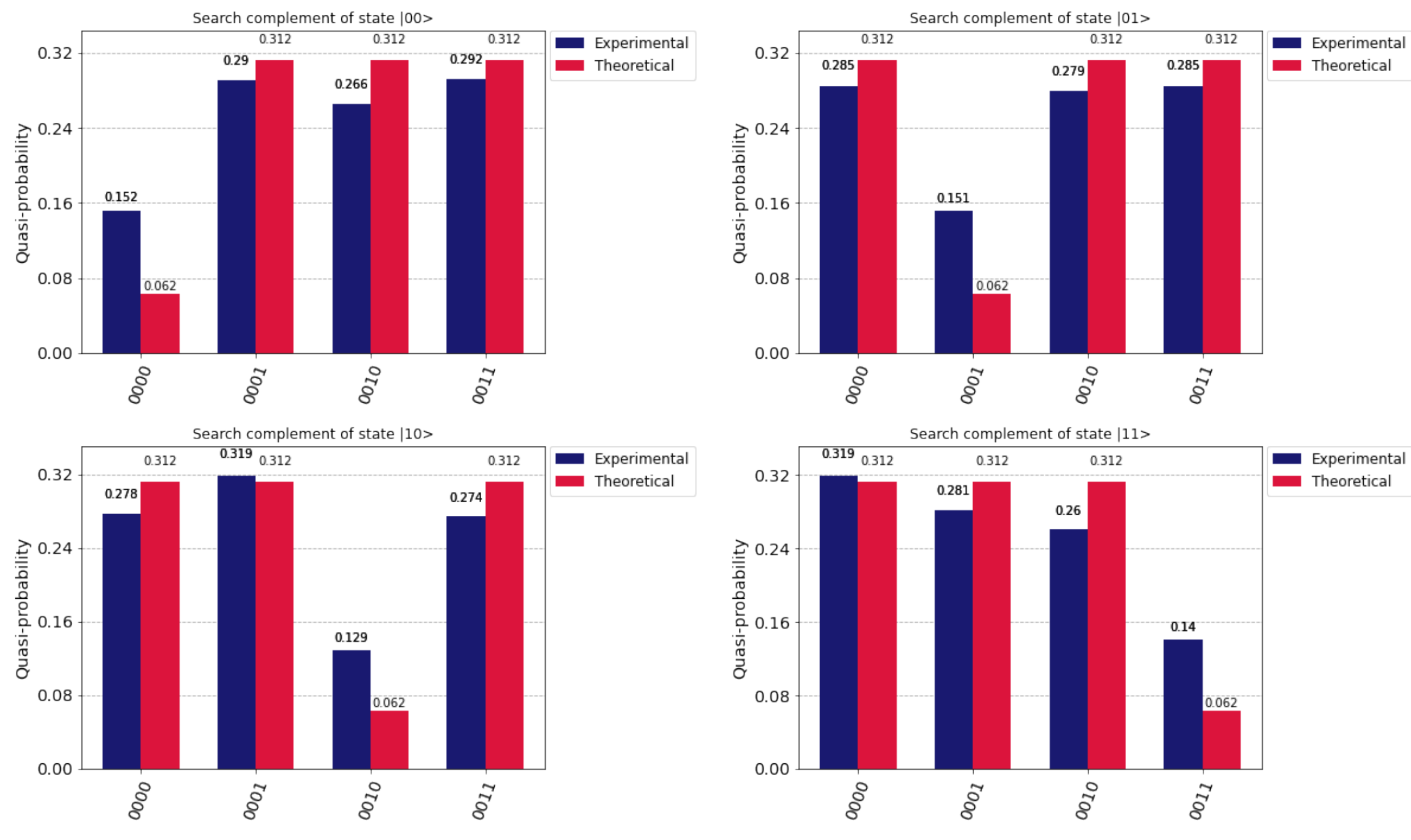}
}
    \caption{Theoretical (right bars) and experimental (left bars) probability distributions after running the search complement algorithm on $ibmq\_manila$ using 20000 shots. The top right, top left, bottom right, and bottom left graphs use the states $|00\rangle$, $|01\rangle$, $|10\rangle$ and $|11\rangle$ as target states, respectively. The theoretical probabilities were obtained from the probability matrices shown in Fig. \ref{search_complement_probability_matrix}.}
    \label{prob_all_targets}
\end{figure}

As explained in Section \ref{section_whole_search_circuit}, for the search complement circuit we must modify the sequence of white and black controls of the perturbed coin operator to vary the target node, and this can be done by adding \textit{CNOT} gates to both sides of the black control we want to negate. Following this rationale, in Fig. \ref{prob_all_targets} we show the execution of the search complement algorithm for all possible target states using the quantum processor \textit{ibmq\_manila}, given that this was the quantum processor that provided the best results among all processors publicly available in IBM Quantum in \cite{wbva202301}. The calibration parameters of the quantum processor \textit{ibmq\_manila} at the running time is shown in Table \ref{imbq_manila_cal} in Appendix \ref{Calibration Data}.

To measure the resemblance of the experimental and the theoretical probability 
distributions, we make use of the statistical distance between two distributions defined as $\ell_1(P,Q) = \frac{1}{2}\sum\limits_{\forall i}|P(i)-Q(i)|$. Table \ref{l1_distance_table_search} displays the $\ell_1$ distances for each run, where the shorter the distance the better the performance of the experimental distributions.

\begin{table}[h!]
\caption{$\ell_{1}$ distances of the comparison between theoretical and experimental distributions of the execution of the search complement algorithm for all possible target states. The shorter the $\ell_{1}$ distance, the closer the resemblance between distributions. The theoretical and experimental probabilities of each state used to calculate $\ell_{1}$ can be seen in Fig. \ref{prob_all_targets}.} 
\label{l1_distance_table_search} 
\centering 
\begin{tabular}{c c c c c} 
\hline\hline 
Target state & Search complement \\ [0.5ex] 
\hline 
$|00\rangle$ & 0.0895\\
$|01\rangle$ & 0.0889\\
$|10\rangle$ & 0.0729\\
$|11\rangle$ & 0.0841\\ [1ex] 
\hline 
\end{tabular}
\end{table}

\section{Conclusion}\label{conclusion_algo_section}

In this work we propose a modified version of the quantum walk-based search algorithm developed by Shenvi, Kempe and Whaley in \cite{shenvi_kempe_whaley_2003}, usually referred to as the SKW algorithm. The SKW algorithm works by increasing the probability of measuring the target state of the search after the repeated application of an evolution operator to a bipartite quantum system. Our modification to the SKW algorithm leads to the opposite behaviour, i.e. we are able to decrease the probability of measuring the target state of the search. In other words, we obtain the \textit{search complement}. Furthermore, it takes only one application of the evolution operator.

The search complement algorithm, is modelled as a one-step Unitary Coined Discrete-Time Quantum Walk (UCDTQW) on a $2^n$-complete graph, $\mathcal{K}_{2^n}$. The UCDTQW uses a perturbed evolution operator that consists of shift operator associated to $\mathcal{K}_{2^n}$ and a perturbed coin which biases the UCDTQW towards all nodes except the target node, creating a "probability leak out" of the target node in $\mathcal{K}_{2^n}$. The probability of measuring the target state is $p_t = \frac{1}{4^n}$ and the probability of measuring any non-target state is $p_{k\neq t} = \frac{1}{2^n} + \frac{1}{4^n}$, where $n$ is the number of qubits of the position register we use to run the algorithm. Given that $p_t$ decreases exponentially faster than $p_{k\neq t}$ when increasing the number of qubits $n$, the functioning of our algorithm relies on increasing the qubits of the system to decrease the probability of measuring the target state. Conceptually, this means we increase the number of nodes in the graph where the one-step UCDTQW takes place, splitting the probability over a larger number of non-target nodes, thus decreasing the probability of the target node.

To analyze the behavior of the algorithm we used the theoretical framework developed in \cite{wbva202302}, where it is proposed that any bipartite quantum operator is associated to a directed multigraph, which is extremely useful for visualization purposes. In this way we were able to see the directed multigraph associated to low-dimensional cases of the evolution operator of the system, which helped us notice that from the point of view of UCDTQWs the algorithm works because there are less arcs pointing towards the target node. Furthermore, we introduced the concept of the probability matrix of a bipartite quantum operator, which is a row-block matrix whose columns are the probability distributions associated to all possible initial states of the form $\ket{i}\otimes\ket{j}$, which helped us understand the behavior of different bipartite evolution operators studied in this work. Additionally, we provided an explanation of the algorithm based on the quantum circuit model of computation, which emphasizes that superposition of quantum states is the main property that makes the algorithm work.

Next, we performed simulations of the search complement algorithm on $\mathcal{K}_4$, $\mathcal{K}_8$ and $\mathcal{K}_{64}$. For $\mathcal{K}_4$ the probability of measuring the target state was $6.2\%$, which is $\thicksim 5$ times smaller than then probability of measuring non-target states ($31.3\%$). For $\mathcal{K}_8$ the probability of measuring the target state is $1.6\%$, which is $\thicksim 8.8$ times smaller than the probability of measuring non-target states ($14.1\%$). Finally, for $\mathcal{K}_{64}$ the probability of getting the state corresponding to node 1, becomes $0.02\%$, which is $\thicksim 80$ times smaller than then probability of measuring non-target states ($1.6\%$). That is, with 12 qubits we were able to reach a probability of nearly zero of measuring the target state.

Finally, we implemented the search complement algorithm on the graph $\mathcal{K}_4$ and ran it four times, each time having a different target node. We ran the quantum circuits on \textit{ibmq\_manila}, a quantum processor made publicly available by IBM quantum, given that \textit{ibmq\_manila} provided the best experimental results for a UCDTQW on the graph $\mathcal{K}_4$ in \cite{wbva202301}. We used the $\ell_1$ distance to compare experimental and theoretical distributions and obtained $\ell_1$ distances $\le 0.0895$ for all quantum circuits.

Similar to the modification we did to the SKW algorithm, in \cite{search_complement_grover} a modification to Grover's algorithm was done to obtain search complement, decreasing the depth of the quantum circuit. The modified circuit was then used as a subroutine for the QAOA, making a better job at minimizing the energy of the system. This suggests that the search complement algorithm based UCDTQW also has the potential to be a subroutine for the QAOA algorithm or other quantum algorithms, giving it a practical use.

In summary, we made use of the directed multigraph framework developed in \cite{wbva202302} along with the probability matrix of a quantum operator, introduced in this work, to analyze and modify the SKW search algorithm, obtaining a search complement algorithm. The efficient implementation of the original search algorithm proposed in \cite{shenvi_kempe_whaley_2003} in a general-purpose quantum computer is still an open problem, however we make progress in the filed of quantum walks by implementing a modified version of the algorithm in a general-purpose quantum computer for the first time. Thus providing evidence that the directed multigraph framework developed in \cite{wbva202302} to interpret UCDTQWs and the probability matrix of an evolution operator have the potential to be useful tools for the analysis and creation of new quantum algorithms.

\bibliographystyle{unsrt}
\bibliography{references}

\section*{Acknowledgements}

The three authors acknowledge the financial support provided by Tecnologico de Monterrey, Escuela de Ingenieria y Ciencias and Consejo Nacional de Ciencia y Tecnología (CONACyT). SEVA acknowledges the support of CONACyT-SNI [SNI number 41594]. SEVA and AWB acknowledge the unconditional support of their families.

\appendix

\section{Formal Proof of the Search Complement Algorithm}
\label{Formal Proof of the Search Complement Algorithm}
In this section we provide a formal proof of the search complement algorithm.

To begin with, in Eq. \eqref{fully_controlled_gate_eq} we provide the analytical form of an arbitrary fully controlled $U$ gate with $n$ top control qubits and $m$ bottom target qubits
\begin{equation}
    \mathcal{C}^n(U) = U \otimes |t\rangle \langle t| + I_2^{\otimes m}\otimes I_2^{\otimes n}-I_2^{\otimes m}\otimes|t\rangle\langle t|
    \label{fully_controlled_gate_eq}
\end{equation}
$|t\rangle \in \mathbf{C}^n$ is the state that activates the target gate, $U$, and $I_2$ is the $2 \times 2$ identity. Notice that $\mathcal{C}^n$ acts as the identity whenever it is applied on a state $|i\rangle$ that is different from $|t\rangle$, but when the state is $|i\rangle = |t\rangle$, then the first and last term in Eq. \eqref{fully_controlled_gate_eq} are activated, and the result is $U|t\rangle$. 

The analytical form of the shift operator of the \textit{CNOT} model to perform a UCDTQW on $\mathcal{K}_{2^n}$ is

\begin{equation}
    S = \sum_{i=0}^{2^m-1} \sum_{j=0}^{2^n-1} \ket{i}\bra{i}\otimes \ket{j\oplus i}\bra{j}
\end{equation}
Once both shift and coin operators of the system are defined, let us obtain a reduced expression for the operator of the search complement algorithm, $\mathcal{U} = S C'(I^{\otimes n}_2\otimes H^{\otimes n})$, where $n$ is the number of qubits of both position and coin registers.

\begin{align}
    \mathcal{U} &= 
    \Big[\sum_{i=0}^{2^m-1} \sum_{j=0}^{2^n-1} \ket{i}\bra{i}\otimes \ket{j\oplus i}\bra{j}\Big]
    \Big[H^{\otimes m} \otimes \ket{t}\bra{t} + I_m\otimes I_m  \nonumber - I_m \otimes \ket{t}\bra{t}\Big]\Big[I^{\otimes m}_2\otimes H^{\otimes n}\Big]\\
    \nonumber\\
    \mathcal{U} &= \sum_{i=0}^{2^m-1} \sum_{j=0}^{2^n-1} \ket{i}\bra{i}H^{\otimes m} \otimes \ket{j\oplus i}\bra{j}\ket{t}\bra{t}H^{\otimes n}\nonumber\\ 
    &\hspace{0cm}+ \sum_{i=0}^{2^m-1} \sum_{j=0}^{2^n-1} \ket{i}\bra{i}\otimes \ket{j\oplus i}\bra{j} H^{\otimes n} \nonumber\\
    &\hspace{0cm}- \sum_{i=0}^{2^m-1} \sum_{j=0}^{2^n-1} \ket{i}\bra{i}\otimes  \ket{j\oplus i}\bra{j}\ket{t}\bra{t} H^{\otimes n}
\end{align}
Using the property $\bra{j}\ket{t} = \delta_{jt}$, $\mathcal{U}$ simplifies as follows

\begin{align}
    \mathcal{U}  &= \sum_{i=0}^{2^m-1} \sum_{j=0}^{2^n-1} \ket{i}\bra{i}H^{\otimes m} \otimes \ket{j\oplus i}\delta_{jt}\bra{t}H^{\otimes n}\nonumber\\ 
    &\hspace{0cm}+ \sum_{i=0}^{2^m-1} \sum_{j=0}^{2^n-1} \ket{i}\bra{i}\otimes \ket{j\oplus i}\bra{j} H^{\otimes n} \nonumber\\
    &\hspace{0cm}- \sum_{i=0}^{2^m-1} \sum_{j=0}^{2^n-1} \ket{i}\bra{i}\otimes \ket{j\oplus i}\delta_{jt}\bra{t} H^{\otimes n}
\end{align}

\begin{align}
    \Rightarrow \mathcal{U} &= \sum_{i=0}^{2^m-1} \ket{i}\bra{i}H^{\otimes m} \otimes \ket{t\oplus i}\bra{t}H^{\otimes n}\nonumber\\ 
    &\hspace{0cm}+ \sum_{i=0}^{2^m-1} \sum_{j=0}^{2^n-1} \ket{i}\bra{i}\otimes \ket{j\oplus i}\bra{j} H^{\otimes n}\nonumber \\
    &\hspace{0cm}- \sum_{i=0}^{2^m-1} \ket{i}\bra{i}\otimes\ket{t\oplus i}\bra{t} H^{\otimes n}
\end{align}
Finally, we get
\begin{align}
\mathcal{U}&=\sum_{i=0}^{2^m-1}\ket{i}\bra{i}\left(H^{\otimes m}-I_m\right)\otimes \ket{t\oplus i}\bra{t}H^{\otimes n}\nonumber\\ 
&\hspace{0cm}+ \sum_{i=0}^{2^m-1} \sum_{j=0}^{2^n-1} \ket{i}\bra{i}\otimes \ket{j\oplus i}\bra{j} H^{\otimes n}
\label{simp_ev_op_complement}
\end{align}
Next, we use Eq. \eqref{simp_ev_op_complement} to calculate the first step of the quantum walk. Consider the initial state of the quantum walk to be
\begin{equation}
    \ket{\psi_0} = \ket{r}\otimes \ket{s}
\end{equation}
Then
\begin{align}
   \ket{\psi_1}&= \mathcal{U}\ket{\psi_0}\\
   \ket{\psi_1}&= \sum_{i=0}^{2^m-1}\ket{i}\bra{i}\left(H^{\otimes m}-I_m\right)\ket{r}\otimes \ket{t\oplus i}\bra{t}H^{\otimes n}\ket{s}\nonumber\\
   &\hspace{0cm}+ \sum_{i=0}^{2^m-1} \sum_{j=0}^{2^n-1} \ket{i}\bra{i}\ket{r}\otimes \ket{j\oplus i}\bra{j} H^{\otimes n}\ket{s}\\
   \ket{\psi_1}&= \sum_{i=0}^{2^m-1}\ket{i}\bra{i}H^{\otimes m}\ket{r}\otimes \ket{t\oplus i}\bra{t}H^{\otimes n}\ket{s}\nonumber\\
   &\hspace{0cm}-\sum_{i=0}^{2^m-1}\ket{i}\bra{i}\ket{r}\otimes \ket{t\oplus i}\bra{t}H^{\otimes n}\ket{s}\nonumber\\
   &\hspace{0cm}+ \sum_{i=0}^{2^m-1} \sum_{j=0}^{2^n-1} \ket{i}\bra{i}\ket{r}\otimes \ket{j\oplus i}\bra{j} H^{\otimes n}\ket{s}\\
   \ket{\psi_1}&= \sum_{i=0}^{2^m-1}\ket{i}\bra{i}H^{\otimes m}\ket{r}\otimes \ket{t\oplus i}\bra{t}H^{\otimes n}\ket{s}\nonumber\\
   &\hspace{0cm}-\sum_{i=0}^{2^m-1}\ket{i}\delta_{ir}\otimes \ket{t\oplus i}\bra{t}H^{\otimes n}\ket{s}\nonumber\\
   &\hspace{0cm}+ \sum_{i=0}^{2^m-1} \sum_{j=0}^{2^n-1} \ket{i}\delta_{ir}\otimes \ket{j\oplus i}\bra{j} H^{\otimes n}\ket{s}
\end{align}
Thus, the final expression for $\ket{\psi_1}$ is given by Eq. \eqref{eq:firststep}.

\begin{align}
    \ket{\psi_1} &= \sum_{i=0}^{2^m-1}\bra{i}H^{\otimes m}\ket{r}\bra{t}H^{\otimes n}\ket{s}\Big(\ket{i}\otimes \ket{t\oplus i}\Big) \nonumber\\
   &\hspace{0cm}-\bra{t}H^{\otimes n}\ket{s} \Big(\ket{r}\otimes \ket{t\oplus r}\Big) \nonumber\\
   &\hspace{0cm}+ \sum_{j=0}^{2^n-1} \bra{j} H^{\otimes n}\ket{s}\Big(\ket{r}\otimes \ket{j\oplus r}\Big)\label{eq:firststep}
\end{align}

Now, let us get the probability of finding the walker at an arbitrary position $\ket{k}$ when measured. For this, let us apply the measurement operator $M_k = I_m \otimes \ket{k}\bra{k}$ to state $\ket{\psi_1}$.

\begin{align}
    M_k \ket{\psi_1} &=  \sum_{i=0}^{2^m-1}\bra{i}H^{\otimes m}\ket{r}\bra{t}H^{\otimes n}\ket{s}\bra{k}\ket{t\oplus i}\Big(\ket{i}\otimes \ket{k}\Big)\nonumber\\
   &\hspace{1cm}-\bra{t}H^{\otimes n}\ket{s} \bra{k}\ket{t\oplus r}\Big(\ket{r}\otimes \ket{k}\Big)\nonumber \\
   &\hspace{1cm}+\sum_{j=0}^{2^n-1} \bra{j} H^{\otimes n}\ket{s} \bra{k}\ket{j\oplus r} \Big(\ket{r}\otimes \ket{k}\Big)
\end{align}

\begin{align}
    \Rightarrow M_k \ket{\psi_1} &=  \sum_{i=0}^{2^m-1}\bra{i}H^{\otimes m}\ket{r}\bra{t}H^{\otimes n}\ket{s}\delta_{k,t\oplus i}\Big(\ket{i}\otimes \ket{k}\Big)\nonumber\\
   &\hspace{0cm}-\bra{t}H^{\otimes n}\ket{s} \delta_{k,t\oplus r}\Big(\ket{r}\otimes \ket{k}\Big) \nonumber\\
   &\hspace{0cm}+\sum_{j=0}^{2^n-1} \bra{j} H^{\otimes n}\ket{s} \delta_{k,j\oplus r}\Big(\ket{r}\otimes \ket{k}\Big)\label{eq:deltas}    
\end{align}

Now, in order to contract the Kronecker deltas in the last expression and to facilitate calculations, consider two integers $u$ and $v$ such that
\begin{align}
    k &= t\oplus u\\
    k&= v\oplus r
\end{align}
This allows us to turn Eq. \eqref{eq:deltas} into
\begin{align}
    M_k \ket{\psi_1}&= \bra{u}H^{\otimes m}\ket{r}\bra{t}H^{\otimes n}\ket{s}\Big(\ket{u}\otimes \ket{k}\Big)\\
   &\hspace{0cm}+\big(\bra{v} H^{\otimes n}\ket{s}-\delta_{k,t\oplus r}\bra{t}H^{\otimes n}\ket{s} \big)\Big(\ket{r}\otimes \ket{k}\Big) \nonumber
\end{align}
The probability is then given by the squared modulus of this expression, i.e.

\begin{align}
    \Rightarrow|M_k \ket{\psi_1}|^2 &= \big(\bra{u}H^{\otimes m}\ket{r}\big)^2 \big(\bra{t}H^{\otimes n}\ket{s}\big)^2 \\
    &\hspace{0cm}+ \big(\bra{v} H^{\otimes n}\ket{s}-\delta_{k,t\oplus r}\bra{t}H^{\otimes n}\ket{s} \big)^2\nonumber\\
    \nonumber\\
\end{align}

\begin{align}
    |M_k \ket{\psi_1}|^2 &= \big(\bra{u}H^{\otimes m}\ket{r}\big)^2 \big(\bra{t}H^{\otimes n}\ket{s}\big)^2 \label{eq:lastone}\\
    &\hspace{0cm}+ \big(\bra{v} H^{\otimes n}\ket{s}\big)^2+\delta_{k,t\oplus r}\big(\bra{t}H^{\otimes n}\ket{s}\big)^2\nonumber\\
    &\hspace{0cm}-\delta_{k,t\oplus r}\bra{t}H^{\otimes n}\ket{s}\bra{v} H^{\otimes n}\ket{s}\nonumber    
\end{align}

Notice then that $\bra{a}H^{\otimes m}\ket{b}$ is the $(a,b)$-th entry in the $m$ dimensional Hadamard matrix. That is,
\begin{equation}
    \bra{a}H^{\otimes m}\ket{b} = \frac{(-1)^{a\cdot b}}{\sqrt{2^m}}
\end{equation}
where $a\cdot b$ is the \textit{binary dot product} or the number $1$ bits $a$ and $b$ have in common in their binary representations. Notice then that the value of $a\cdot b$ does not matter when obtaining the square of the entry
\begin{equation}
    \big(\bra{a}H^{\otimes m}\ket{b}\big)^2 = \frac{1}{2^m}
\end{equation}
This turns Eq. \eqref{eq:lastone} into
\begin{align}
    p_k &= \Big(\frac{1}{2^m}\Big) \Big(\frac{1}{2^n}\Big)+ \frac{1}{2^n}+\frac{\delta_{k,t\oplus r}}{2^{n}}-\delta_{k,t\oplus r}\bra{t}H^{\otimes n}\ket{s}\bra{v} H^{\otimes n}\ket{s}
\end{align}
Which is the general expression for the probability of measuring a node using the search complement algorithm.

Notice that we still have one Kronecker delta left, which leaves us with two cases. 

\begin{enumerate}
    \item Case 1: When $k \neq t\oplus r$

The probability of measuring the state associated to node $k$ is

\begin{align}
    p(v_k) = \frac{1}{2^{m+n}}+\frac{1}{2^n}
\end{align}

    \item Case 2: When $k=t\oplus r$

The probability of measuring the state associated to node $k$ is

\begin{align}
    p(v_k) &= \frac{1}{2^{m+n}}+ \frac{1}{2^n}+\frac{1}{2^{n}}-2\bra{t}H^{\otimes n}\ket{s}\bra{v} H^{\otimes n}\ket{s}
\end{align}    
    
\end{enumerate}

Moreover, as $v$ is defined by $k = v \oplus r$ and $k=t\oplus r$, we must have $v=t$, and thus:
\begin{align}
    p(v_k) &= \frac{1}{2^{m+n}}+ \frac{2}{2^n}-2\big(\bra{t}H^{\otimes n}\ket{s}\big)^2\nonumber\\
    &=  \frac{1}{2^{m+n}}+ \frac{2}{2^n}-\frac{2}{2^n}\nonumber\\
    &=  \frac{1}{2^{m+n}}\nonumber
\end{align}

Lastly, given that the coin state is represented with the same number of bits as the position state in $\ket{\psi}$, that is $m=n$, we get the following final expression for the probability distribution of a quantum walker:

\begin{align}
    p(v_k) = \begin{cases}
        \frac{1}{4^{n}} &\text{if $k=t\oplus r$}\\
        \frac{1}{4^n}+\frac{1}{2^n} &\text{if $k \neq t\oplus r$}
    \end{cases}
\label{general_prob_dist_complement_appendix}
\end{align}

In the case where $\ket{r} = \ket{0}$, then the probability of measuring the target state is $p(v_t) = \frac{1}{4^n}$, while the probability of measuring any other state is $p(v_{k \neq t}) = \frac{1}{4^n}+\frac{1}{2^n}$. Thus, the probability of measuring the target state $\ket{t}$ is indeed minimized.

To prove that Eq. \eqref{general_prob_dist_complement_appendix} is indeed a probability distribution we calculate the sum of all probabilities, i.e.

\begin{align}
    \sum_{i=0}^{2^n - 1} p(v_i) &= p(v_{t\oplus r})+\sum_{i\neq t\oplus r} p(v_i) \nonumber
\end{align}

Given that there are $2^n-1$ non-target states, we get 

\begin{align}
    \sum_{i=0}^{2^n - 1} p(v_i) = \frac{1}{4^n}+\Big(2^n - 1\Big)\Big( \frac{1}{4^n}+\frac{1}{2^n} \Big)=\frac{1}{4^n}+\Big( \frac{1}{2^n}+1 \Big)-\Big( \frac{1}{4^n}+\frac{1}{2^n} \Big)\nonumber
\end{align}

\begin{align}
    \therefore \sum_{i=0}^{2^n - 1} p(v_i) = 1 \nonumber
\end{align}

\newpage

\section{Calibration Data}
\label{Calibration Data}

In this section we provide the calibration parameters of the quantum processor imbq\textunderscore manila at the time of running the search complement algorithm to obtain the experimental data for Figure 14 and Table 1 in the main text.

\begin{table}[h!]
\caption{Calibration of quantum processor imbq\textunderscore manila when executing the search complement algorithm. The results of the algorithm are shown in Figure 14 and Table 1 of the main text.} 
\label{imbq_manila_cal} 
\centering 
\scalebox{0.65}{
\begin{tabular}{c c c c c c} 
\hline\hline 
Qubit/Parameter 
& $Q_0$ 
& $Q_1$ 
& $Q_2$ 
& $Q_3$ 
& $Q_4$\\
\hline 
$T_1$($\mu$s) 
& 149.09
& 203.7
& 161.46
& 203.01
& 108.82
\\ 
$T_2$ ($\mu$s) 
& 90.93
& 87.48
& 22.82
& 58.99
& 17.28
\\
Frequency (GHz) 
& 4.963
& 4.838
& 5.037
& 4.951
& 5.066
\\
Anhamonicity (GHz) 
& -0.34335
& -0.34621
& -0.34366
& -0.34355
& -0.34211
\\ 
Readout assignment error ($10^{-2}$)
& 2.05
& 1.71
& 1.99
& 1.95
& 2.09
\\
Prob meas0 prep1 
& 0.0306
& 0.152
& 0.03
& 0.0256
& 0.0316
\\
Prob meas1 prep0
& 0.0104
& 0.1902
& 0.0098
& 0.0134
& 0.0102
\\
Readout length (ns)
& 5351.111
& 5351.111
& 5351.111
& 5351.111
& 5351.111
\\
ID error ($10^{-4}$)
& 1.83
& 6.84
& 3.18
& 2.09
& 4.19
\\
$\sqrt{x}$ error ($10^{-4}$)
& 1.83
& 6.84
& 3.18
& 2.09
& 4.19
\\
Single-qubit Pauli-X error ($10^{-4}$)
& 1.83
& 6.84
& 3.18
& 2.09
& 4.19
\\
CNOT error ($10^{-2}$)
& 0-1:1.335
& 1-2:1.715; 1-0:1.335
& 2-3:1.126; 2-1:1.715
& 3-4:0.5390; 3-2:1.126
& 4-3:0.5390
\\
Gate time (ns)
& 0-1:277.333
& 1-2:469.333; 1-0:312.889
& 2-3:355.556; 2-1:504.889
& 3-4:334.222; 3-2:391.111
& 4-3:298.667
\\[1ex] 
\hline 
\end{tabular}
}
\end{table} 

\end{document}